\documentclass[aps,twocolumn,noshowpacs,superscriptaddress,floatfix]{revtex4}

\usepackage[utf8]{inputenc}
\usepackage{braket}
\usepackage{amsmath,amssymb,graphicx}
\usepackage{caption}
\usepackage{subcaption}
\usepackage{qcircuit}
\usepackage{xcolor}
\usepackage{verbatim}
\usepackage[normalem]{ulem}
\usepackage{graphicx}
\usepackage[rightcaption]{sidecap}

\newcommand{\norm}[1]{\left\lVert#1\right\rVert}
\newcommand{\R}{\mathbb{R}}

\def\ket#1{\mathinner{|{#1}\rangle}}

\usepackage{float}
\usepackage{multirow}

\begin{document}

\title{Medical image classification via quantum neural networks}

\author{Natansh Mathur}
\affiliation{QC Ware, Palo Alto, USA and Paris, France}
\affiliation{Indian Institute of Technology Roorkee, India}

\author{Jonas Landman}
\affiliation{QC Ware, Palo Alto, USA and Paris, France}
\affiliation{IRIF, CNRS - University of Paris, France}

\author{Yun Yvonna Li}
\affiliation{F. Hoffmann La Roche AG}

\author{Martin Strahm}
\affiliation{F. Hoffmann La Roche AG}

\author{Skander Kazdaghli}
\affiliation{QC Ware, Palo Alto, USA and Paris, France}

\author{Anupam Prakash}
\affiliation{QC Ware, Palo Alto, USA and Paris, France}

\author{Iordanis Kerenidis}
\affiliation{QC Ware, Palo Alto, USA and Paris, France}
\affiliation{IRIF, CNRS - University of Paris, France}

\date{September 2021}

\begin{abstract}

Machine Learning provides powerful tools for a variety of applications, including disease diagnosis through medical image classification. In recent years, quantum machine learning techniques have been put forward as a way to potentially enhance performance in machine learning applications, both through quantum algorithms for linear algebra and quantum neural networks. 
In this work, we study two different quantum neural network techniques for medical image classification: first by employing quantum circuits in training of classical neural networks, and second, by designing and training quantum orthogonal neural networks. We benchmark our techniques on two different imaging modalities, retinal color fundus images and chest X-rays. The results show the promises of such techniques and the limitations of current quantum hardware.

\end{abstract}

\maketitle

\noindent\textbf{Note: The final published version of this work, combined with \cite{klm21}, appears as \cite{Landman2022quantummethods}.}

\section{Introduction}

Medical image classification is an important application of machine learning in the healthcare domain. As deep learning algorithms continue to achieve accuracy levels on par with humans, numerous efforts are taken to automate diagnostic, or even prognostic tasks \cite{Benjamens}. Despite the success of artificial neural networks to classify, segment, and perform other image related tasks, complexity of such models and the cost of training continues to increase. One of the latest visual transformer architecture trained on imageNet boasts of more than 2 billion parameters, trained used more than 10 000 TPU core-days \cite{Zhai}. This begs the question whether alternative technologies can provide better solutions in terms of scalability. 

Recently with the advent of the first small-scale noisy quantum computers, the study of quantum machine learning, Quantum Neural Network (QNN) techniques in particular, has received a lot of attention. While current quantum hardware is far from being powerful enough to compete with classical machine learning algorithms, the first proof of concept demonstrations of interesting quantum machine learning algorithms have started to appear \cite{QNN2018,CNN2019,Image2020, Semisupervised2020,Polyadic2020,Dressed2020,Supervised2018,Hierarchical2018, NearestCentroid2021,Kiani20}. 
In \cite{Supervised2018}, the authors use parametrized quantum circuits where parameters are trained through stochastic gradient descent, and results on synthetic data are presented in comparison with results using Support Vector Machines classification.
In \cite{Hierarchical2018} the authors outline different classification methods based on different quantum parametrized circuits, where the classical data is encoded as separable qubits. A 4-qubit hardware experiment for a binary classification task between two of the IRIS dataset classes was performed on an IBM machine with high accuracy. 
In \cite{Polyadic2020}, the authors performed 2-qubit experiments on IBM machines using the IRIS dataset. 
Other near term implementations include small fully connected neural networks \cite{beer2020training,abbas2021power}, convolutional neural networks \cite{cong2019quantum}, and generative models \cite{xiaodi_neurips,coyle2020born}. They all rely on the parameter shift rule \cite{mitarai2018quantum} method to compute the gradients of the gate's parameters, and give encouraging results on small scale experiments. More details on noisy intermediate scale quantum algorithms can also be found in \cite{bharti2021noisy}.

While such first demonstrations are promising, we neither have solid theoretical evidence that such quantum architectures can be easily trained to provide high accuracy, nor can we perform large enough simulations to get convincing practical evidence of their performance, since we do not have large enough quantum hardware and classical simulations usually incur an exponential overhead. 
For example, architectures that use quantum circuits of constant depth and gates connecting only neighboring qubits are suitable for implementation on near-term quantum computers, but cannot act as a fully connected neural networks since each input qubit can only affect a constant number of output qubits. Further, the time to train such quantum parametrized circuits can be quite large, considering phenomena such as barren plateaus and the fact that designing the architectures, choosing cost-functions and initializing the parameters are far more complex and subtle than one may naively think \cite{QCNNplateaus2020, VQCA2020, Train2020}. Further work is certainly needed to understand the power and limitations of parametrized quantum circuits for machine learning applications. 

Another avenue for quantum supervised learning involves the use of similarity-based techniques to increase speed and accuracy. In \cite{NearestCentroid2021}, a quantum Nearest Centroid classifier was executed on an 11-qubit trapped-ion machine of IonQ, and classification of all ten classes of the MNIST handwritten-digit dataset was performed with an accuracy of $77.5\%$. These quantum similarity-based machine learning methods are interpretable and their performance can be theoretically studied, but may require considerably bigger and better quantum hardware to compete with classical deep learning classification methods.  


Our work is a collaboration between experts in classical machine learning methods for medical image classification and experts in quantum algorithms to advance the state-of-the-art of quantum methods for medical image classification. The hardware demonstration remains a simple proof of concept, which nevertheless marks progress towards unblocking a number of theoretical and practical bottlenecks for quantum machine learning. 

We focus on quantum neural networks for medical image classification and implement two different methods: the first uses quantum circuits to assist the training and inference of classical neural networks by adapting the work in \cite{QNN2020, QCNN2019} to be amenable to current quantum hardware; the second method builds on the recent work on quantum orthogonal neural networks \cite{klm21}, where quantum parametrized circuits are trained to enforce orthogonality of weight matrices. Orthogonality is an inherent property of quantum operations, which are described by unitary matrices, and it has been shown that it can improve performance of deep neural nets and help avoid vanishing or exploding gradients  \cite{nosarzewski2018deep,wang2020orthogonal,jia2019orthogonal}. 

The datasets used to demonstrate the utilities of the two different quantum techniques come from the MedMNIST series \cite{medmnist}.  Two very different imaging modalities, the RetinaMNIST dataset of retina fundus images \cite{retina} and the PneumoniaMNIST dataset of chest X-rays \cite{pneumonia} are chosen. The same techniques can also be easily adapted to be tested on all other datasets from the MedMNIST. 

The hardware experiments both for training the quantum neural networks and forward inference have been performed on the IBM quantum hardware. More specifically, 5-qubit experiments have been performed on three different machines with 5, 7, and 16 qubits, and 9-qubit experiments have been performed in a 16-qubit machine. The IBM quantum hardware is based on superconducting qubits, one of the two most advanced technologies for quantum hardware right now along with trapped ions. We also performed the same training and inference experiments using classical methods as well as simulators of the quantum neural networks for benchmarking the accuracy of the quantum methods.

Our results show quantum neural networks could be used as an alternative to classical neural networks for medical image classification. Results of simulation of the quantum neural networks provide similar accuracy to the classical ones.
From the hardware experiments, we see that for a number of classification tasks the quantum neural networks can be indeed trained to the same level of accuracy as their classical counterparts, while for more difficult tasks, quantum accuracies drop due to the quantum hardware limitations. The MedMNIST datasets provide good benchmarks for tracking the advances in quantum hardware, to act as a proxy for checking how well different hardware perform in classification tasks, and also for testing different classification algorithms and neural network architectures in the literature.

\section{Results}

\subsection{Datasets and pre-processing}

In order to benchmark our quantum neural network techniques for medical images we used datasets from MedMNIST, a collection of 10 pre-processed medical open datasets \cite{medmnist}. The collection has been standardized for classification tasks on lightweight $28 \times 28$ medical images of 10 different imaging modalities. 

In this work we looked specifically at two different datasets. The first is the PneumoniaMNIST \cite{pneumonia}, a dataset of pediatric chest X-ray images. The task is binary-class classification between pneumonia-infected and healthy chest X-rays. The second is the RetinaMNIST, which is based on DeepDRiD \cite{retina}, a dataset of retinal fundus images. The original task is ordinal regression on a 5-level grading of diabetic retinopathy severity, which has been adapted in this work to a binary classification task to distinguish between normal (class 0) and different levels of retinopathy (classes 1 to 4).  

In the PneumoniaMNIST the training set consists of 4708 images (class 0: 1214, class 1: 3494) and the test set has 624 points (234 - 390). In the RetinaMNIST the training set has 1080 points (486 - 594) and the test set 400 (174 - 226). Note that in the RetinaMNIST we have considered the class 0 (normal) versus classes $\{1,2,3,4 \}$ together (different levels of retinopathy).

Though the image dimension of the MedMNIST is small compared to the original images, namely 784 pixels, one cannot load such data on the currently available quantum computers and thus a standard dimensionality reduction pre-processing has been done with Principal Component Analysis to reduce the images to 4 or 8 dimensions. Such a pre-processing indeed may reduce the possible accuracy of both classical and quantum methods but our goal is to benchmark such approaches on current quantum hardware and understand if and when quantum machine learning methods may become competitive.

\subsection{Quantum data-loaders}

Once the classical medical image datasets have been pre-processed we need to find ways of loading the data efficiently into the quantum computer. We use exactly one qubit per feature and outline three different ways of performing what is called a unary amplitude encoding of the classical data points in the following section. 

We use a two-qubit parametrized gate, called Reconfigurable Beam Splitter gate (RBS) in \cite{NearestCentroid2021} and that has appeared also as partial SWAP or fSIM gate, which is defined as 
\begin{equation} \label{RBS}
RBS(\theta) = \left( \begin{array}{cccc}
1 & 0 & 0 & 0 \\
0 & \cos \theta & \sin \theta & 0 \\
0 & -\sin\theta & \cos\theta & 0 \\
0 & 0 & 0 & 1  \end{array} \right)
\end{equation} 

\begin{figure}[h]
    \centering
    \includegraphics[width=100px]{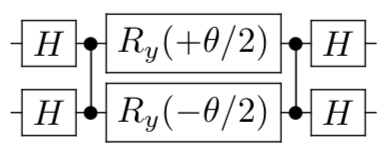}
    \caption{A possible decomposition of the $RBS(\theta)$ gate.}
    \label{fig:RBS_implementation}
\end{figure}

The RBS gate can be easily implemented on hardware through the following decomposition in Fig. \ref{fig:RBS_implementation}, where $H$ is the Hadamard gate, $R_{y}(\theta)$ is a single qubit rotation with angle $\theta$ around the $y$-axis, and the two two-qubit gates represent the CZ gate that flips the sign when both qubits are in state 1.

The first step is a procedure that given access to a classical data point $x = (x_1,x_2,\ldots, x_d) \in \R^d$, pre-processes the classical data efficiently, i.e. spending only $\widetilde{O}(d)$ total time (where we hide logarithmic factors), in order to create a set of parameters $\theta = (\theta_1,\theta_2,\ldots, \theta_{d-1}) \in \R^{d-1}$, that will be the parameters of the $(d-1)$ two-qubit gates we will use in our quantum circuit. In the pre-processing, we also keep track of the norms of the vectors. 

\begin{figure}[!h]
\includegraphics[width=0.45\textwidth]{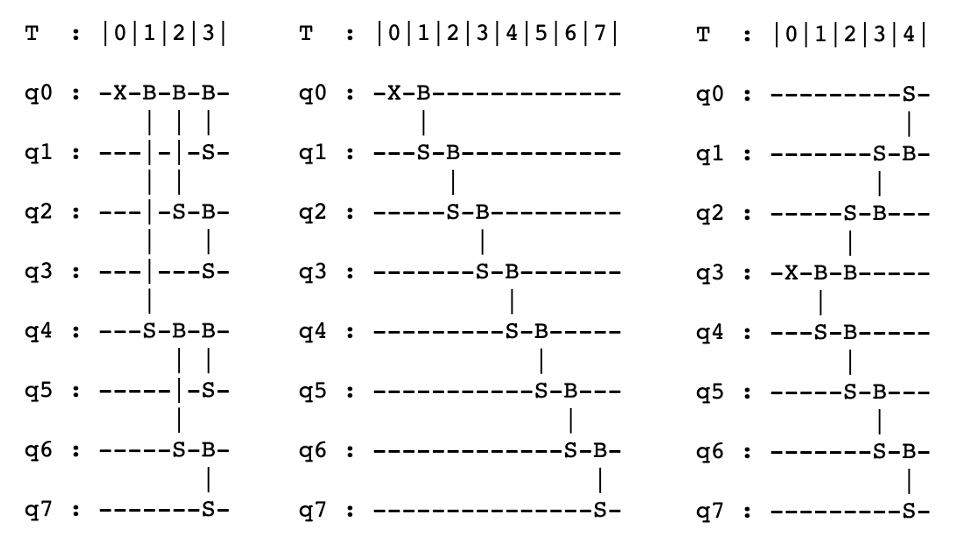}
\caption{The parallel, diagonal, and semidiagonal data loader circuit for an 8-dimensional data point. The X corresponds to the single-qubit Pauli X gate, while the B-S symbols between two qubits correspond to the two-qubit RBS gate (B: first qubit, S: second qubit).}
\label{loaders}
\end{figure}

In the end of the data loader circuits we have created the state
\begin{equation}\label{state}
\ket{x} = \frac{1}{\norm{x}}\sum_{i=1}^{d} x_i \ket{e_i} 
\end{equation}
where the states $\ket{e_i}$ are a unary representation of the numbers $1$ to $d$, using $d$ qubits. Three different types of data loader circuits appear in Figure \ref{loaders}.

The shallowest data loader is a parallel version which loads $d$-dimensional data points using $d$ qubits, $d-1$ RBS gates and circuits of depth only $\log d$ (see first circuit in Figure \ref{loaders}). 


While this loader has the smallest depth, it also requires connectivity beyond the available connectivity on the IBM hardware. For experiments on quantum computers with a small number of qubits, as the ones available nowadays, the difference in the depth between the different data loader circuits is not significant. However as the number of qubits increases, being able to use the parallel data loader can be very beneficial: for example, for a 1024-dimensional data point, the circuit depth can be reduced from 1024 to 10. 

The two data loaders we use in this work have worse asymptotic depth but respect the nearest neighbors connectivity of the IBM hardware.
The first is a simple diagonal unary loader that still uses $d$ qubits and $d-1$ RBS gates between neighboring qubits, but it has circuit depth of $d-1$ (see second circuit in Figure \ref{loaders}). 
The second one is another unary loader whose depth now decreases to $d/2$, and that we refer to as semi-diagonal (see third circuit in Figure \ref{loaders}). 

We provide the details for finding these parameters in the Methods section. 
Such unary amplitude encoding loaders have been used before in different scenarios and the details of how to calculate the set of parameters $\theta$ from a classical data point $x$ appear in \cite{NearestCentroid2021, unary2019}.   

\subsection{Quantum-assisted Neural Networks}

Our first method for exploring QNN is a hybrid approach where the quantum computer is used in the framework of a classical neural network architecture in order to train the neural network and perform forward inference. 

In \cite{QNN2020, QCNN2019}, it was shown how quantum techniques from linear algebra can be used in order to potentially offer speedups in the training of classical neural networks, for both fully connected and convolution nets. These quantum methods are still outside the capabilities of current quantum hardware, so we design and implement in this work Noisy Intermediate Scale Quantum computing (NISQ) friendly subroutines for performing matrix-matrix multiplications through simpler quantum circuits to estimate inner products between two vectors. These circuits have the same behaviour as the more complex quantum linear algebraic procedures, where the quantities that need to be computed can be estimated through sampling a quantum state that encodes these quantities in its amplitudes. This allows us to study how this estimation, instead of an exact computation as is the case in practice classically, affects the performance of the quantum deep learning methods; this also allows us to find ways to mitigate the errors of the quantum hardware.

Important to note here is that simply sampling a quantum state is enough to estimate the squares of the amplitudes of the desired state, which can be enough if we know that the quantity we are interested in is positive. For estimating the inner product of two vectors, which can be either positive or negative, one needs to find efficient ways to estimate the sign of the amplitudes as well.

\begin{figure}[!h]
\includegraphics[width=0.45\textwidth]{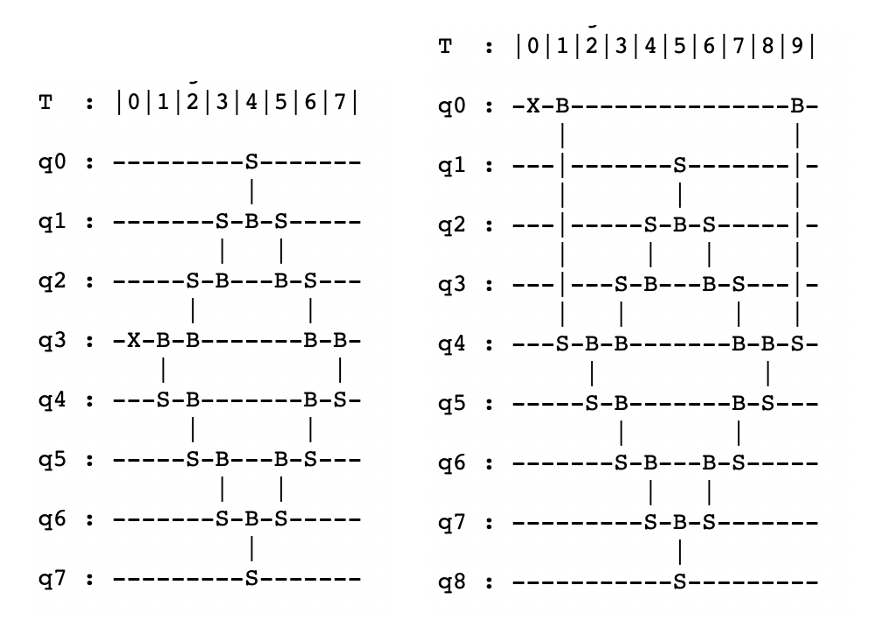}
\caption{The square inner product and the inner product estimation circuits with semi-diagonal data loaders for 8-dimensional data points $x$ and $w$. The RBS gates on the left part of the circuits have parameters that correspond to the first vector $x$, while the RBS gates on the right part of the circuits have parameters that correspond to the second vector $w$.}
\label{fig:IP_circ}
\end{figure}

We start by giving a simpler circuit that can be used for computing the square of the inner product between two normalized vectors $x$ and $w$, as shown in the left circuit in Figure \ref{fig:IP_circ}, which is the same type of circuit used in \cite{NearestCentroid2021}. The first part of the circuit is the data loader for some vector $x$, which in our case will be the normalized data point, and the second part is the adjoint data loader circuit for some vector $w$, which in our case will be a row of the weight matrix. The parameters of the gates in the first part of the circuit are fixed and correspond to the input data points, while the parameters of the gates of the second part are updated as we update the weight matrix through a gradient descent computation. Note that in \cite{NearestCentroid2021} the parallel loader was used, since the experiments were performed in a fully connected trapped ion quantum computer, while here we use the semi-diagonal ones, due to the restricted connectivity of the superconducting hardware.

The final state of this circuit has the form
\begin{equation}
w\cdot x \ket{e_1} + \ket{G}
\end{equation}
where $\ket{G}$ is an unnormalised state orthogonal to $e_1$, in other words whose first qubit is in state $\ket{0}$. This implies that when we measure the first qubit of this quantum state, the probability we get outcome $1$ is exactly $(w\cdot x)^2$, while the outcome is 0 with the remaining probability $1-(w\cdot x)^2$. Thus, if we run the same quantum circuit $N_{shot}$ times and count the number of times the measurement of the first qubit gave outcome 1, we denote it by $N_1$, then we can estimate the square inner product between the two vectors as $N_1/N_{shot}$. A simple Chernoff bound shows that if we want the estimate to be within $\epsilon$ of the correct answer, then we need to take $N_{shot}=O(1/\epsilon^2)$. 

We now need to extend the above circuit in order to be able to estimate the inner product directly and not its square.
While there are different possible ways to do so, we decided on the way shown in the second circuit in Figure \ref{fig:IP_circ}, which adds one extra qubit in the quantum circuit and only a constant number of gates, independent of the dimension of the data points, and it does not need multiple measurements that form a complete operator basis on the Hilbert space of the system.

In short, in order to compute the inner product of two vectors $x$ and $w$, the extra qubit, denoted as $q0$, is initialized with an $X$ gate to $\ket{1}$ and used as a control qubit through an RBS gate between the qubits $q0$ and $q1$ with $\theta=\pi/4$. This way when the qubit $q0$ is 1 nothing happens, and when it is 0, the quantum circuit described above consisting of a quantum data loader for $x$ and the adjoint quantum data loader for $w$ is performed. By performing a final RBS gate with $\theta=\pi/4$ we end up with a final state of the form
\begin{equation}
\left( \frac{1}{2} - \frac{1}{2}w\cdot x \right) \ket{e_1} + \ket{G}
\end{equation}
where the $\ket{G}$ is an unnormalised state orthogonal to $e_1$, in other words whose first qubit is in state $\ket{0}$. Thus, as described above, measuring the first qubit enough times and counting the number of times the outcome is 1, we can estimate the quantity $\left( \frac{1}{2} - \frac{1}{2}w\cdot x \right)^2$ from which, using the fact that $\left( \frac{1}{2} - \frac{1}{2}w\cdot x \right) \geq 0$, we can provide an estimate of $w\cdot x$ directly.
Note also that the connectivity of the new circuit necessitates one qubit with three neighbors while all other qubits can be on a line, which is available on the latest IBM machines. 

To see how the above quantum circuits can assist in the training and inference of classical neural networks: any time there is need for multiplication between data points and weights, instead of classically performing this operation, we can employ the quantum circuit. The advantage these circuits offer is two-fold. First, with the advent of faster quantum computers that also have the possibility to apply gates in parallel (see for example \cite{parallelGates}), one could take advantage of the fact that these quantum circuits require only logarithmic depth (using the parallel loaders) in order to provide a sample from the random variable we are estimating, and thus for larger feature spaces one can expect a speedup in the estimation. 
A second advantage can come from the fact that one can perform training in a different optimization landscape, that of the parameters of the angles of the quantum gates and not the elements of the weight matrices. This can provide different and potentially better models. We will show how to perform such training for the second quantum method in the section below.

\subsection{Quantum Orthogonal Neural Networks}

The second method we use for medical image classification is based on quantum orthogonal neural networks, recently defined in \cite{klm21}. An orthogonal neural network is a neural network where the trained weight matrices are orthogonal matrices, a property that can improve accuracy and avoid vanishing gradients \cite{jia2019orthogonal}. 

The input data is first loaded into the quantum circuit using a diagonal data loader and then a circuit in the form of a pyramid made of RBS gates is applied in order to implement an orthogonal layer of the neural network. 

\begin{figure}[!h]
    \centering
    \includegraphics[width=0.5\textwidth]{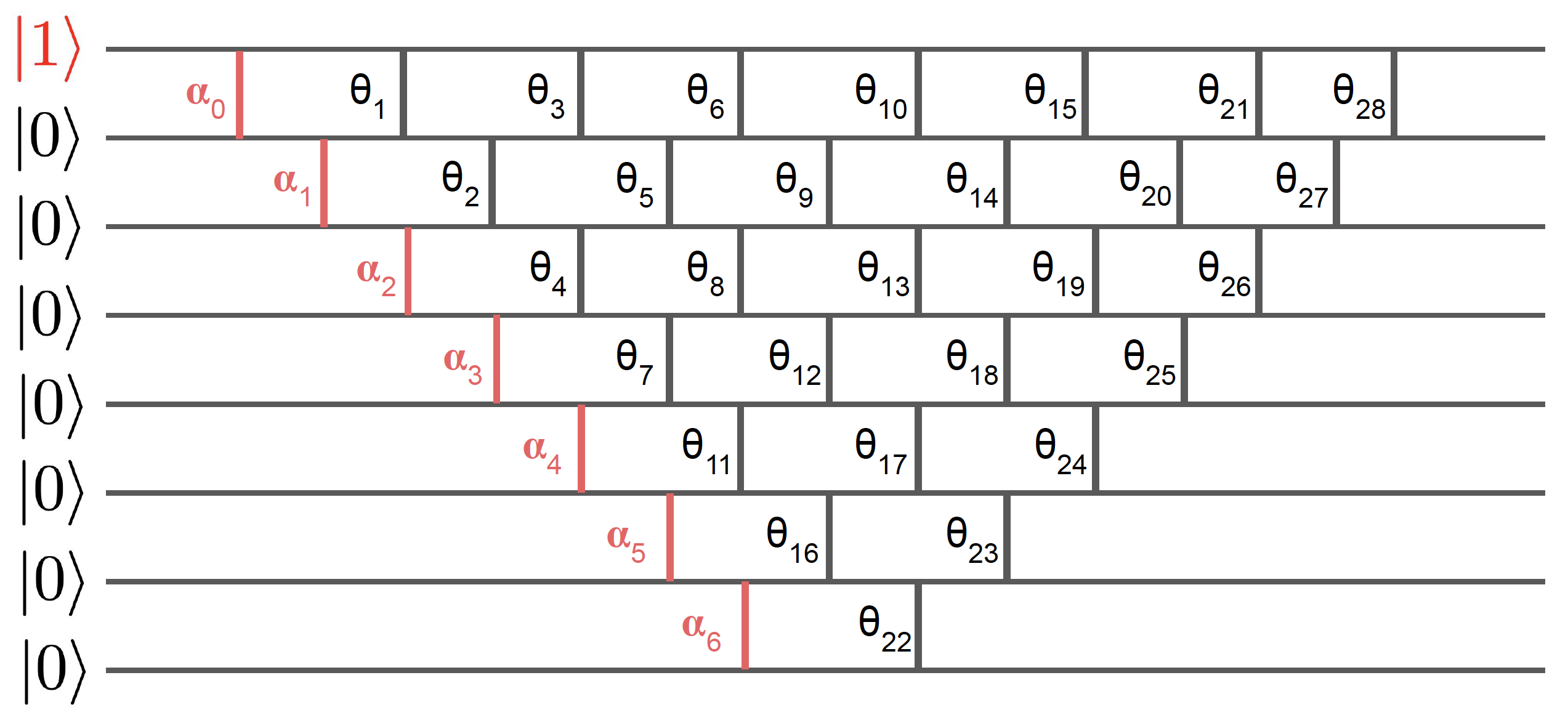}
    \caption{The 8 dimensional diagonal data loader circuit (in red) is efficiently embedded before the quantum pyramid circuit that corresponds to an $8\times 8$ orthogonal layer. The RBS gates are here denoted by a vertical line between the neighboring qubits. The $\alpha$ parameters correspond to the angles of the RBS gates of the diagonal data loader, while the $\theta$ parameters correspond to the angles of the RBS gates of the quantum pyramid circuit \cite{klm21}.}
    \label{fig:data_loader}
\end{figure}

\begin{figure}[!h]
    \centering
    \includegraphics[width=0.35\textwidth]{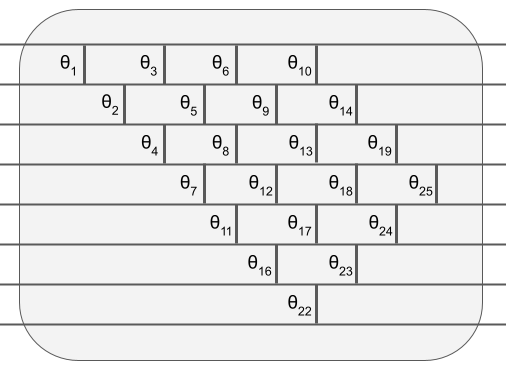}
    \caption{A quantum pyramid circuit that corresponds to an $8\times 4$ orthogonal layer. The parameters $\theta$ correspond to the angles in the RBS gates of the quantum pyramid circuit \cite{klm21}.}
    \label{fig:rectangular}
\end{figure}

The circuit in Figure \ref{fig:data_loader} corresponds to an $8\times 8$ orthogonal layer with a diagonal loader followed by a quantum pyramid circuit. In cases where the number of inputs and outputs are not the same, the pyramid circuit can be curtailed to provide a smaller circuit as shown in Figure \ref{fig:rectangular}.

The important property to note is that the number of parameters of the quantum pyramid circuit corresponding to a neural network layer of size $n \times d$ is $(2n-1-d)\cdot d/2$, exactly the same as the number of degrees of freedom of an orthogonal matrix of dimension $n \times d$.
In fact, there is a one-to-one mapping between any $n\times n$ orthogonal matrix $W$ and the parameters of the quantum gates of the quantum pyramid circuit, so that the matrix $W$ appears as the submatrix of the unitary matrix specified by the quantum pyramid circuit when we restrict to the unary basis vectors. In other words, given an orthogonal weight matrix one can find the corresponding parameters of the quantum gates and vice versa. The mapping holds for the case of $n \times d$ matrices as well. 

If we denote by $x$ the classical data point that is loaded by the diagonal loader, and $W$ the orthogonal weight matrix corresponding to the pyramid circuit with rows $W_j$, then the state that comes out of the circuit in Figure \ref{fig:data_loader} is
\begin{equation}
\sum_{j} W_{j}\cdot x \ket{e_j}
\end{equation}
where we see that the amplitudes correspond to the multiplication of the matrix $W$ with the input $x$.

To be more precise, while the above mentioned pyramid circuit creates the quantum state that corresponds to the multiplication of the orthogonal matrix with the datapoint, in order to get the signs of the coordinates of the resulting vector (and not simply the squared values) one needs to augment the circuit by adding one extra qubit that works as a control qubit, as well as a fixed loader for the uniform vector. The final circuit that we implement on the hardware has the form shown in Figure \ref{final_circuit}.

\begin{figure}[h]
\centering
\begin{subfigure}{.4\textwidth}
  \centering
  \includegraphics[width=\linewidth]{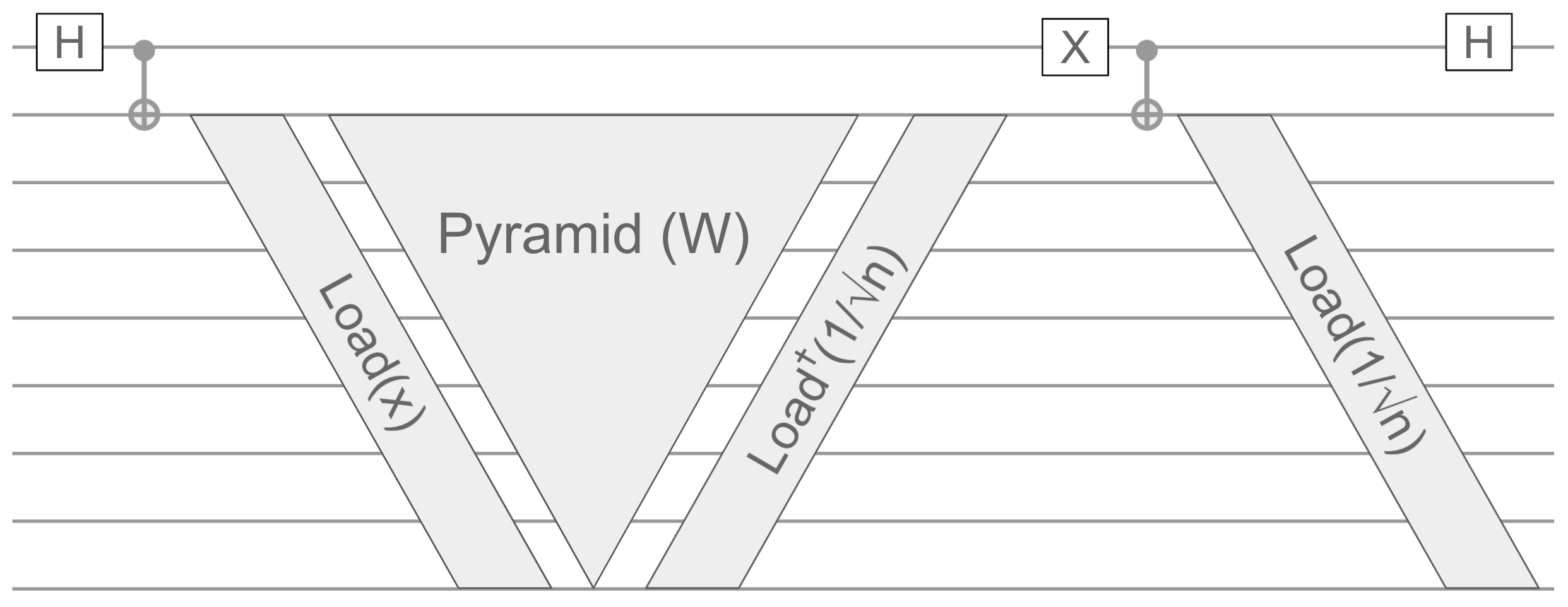}
  \caption{}
  \label{fig:QONNcircuit_rectangular}
\end{subfigure}%
\newline
\begin{subfigure}{.4\textwidth}
  \centering
  \includegraphics[width=\linewidth]{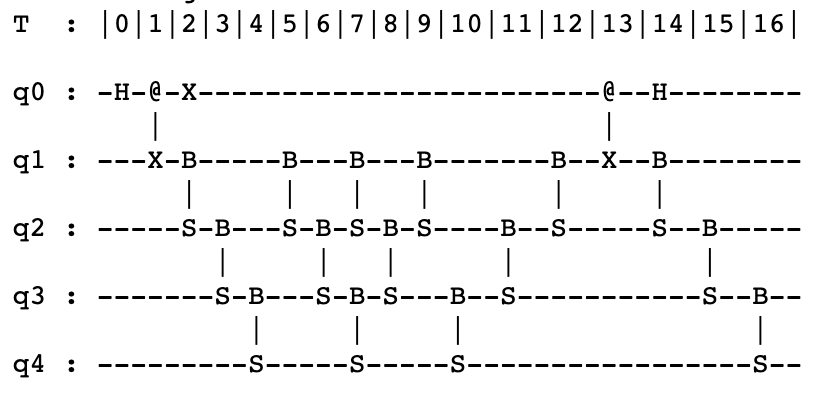}
  \caption{}
  \label{final_circuit}
\end{subfigure}
\caption{ The final quantum circuit for (a) an 8x8 and (b) an 4x4 quantum orthogonal layer, where the symbols @-X correspond to a CNOT gate (@: control qubit, X: target qubit) \cite{klm21}.}
\label{final_circuit}
\end{figure}

The state at the end of this circuit is in fact 
\begin{equation}
    \frac{1}{2}\ket{0}\sum_{j=1}^n \left(W_j x + \frac{1}{\sqrt{n}} \right)\ket{e_j}
    + \frac{1}{2}\ket{1}\sum_{j=1}^n \left(W_j x - \frac{1}{\sqrt{n}} \right)\ket{e_j}
\end{equation}

On this final state, we can see that the difference in the probabilities of measuring the first qubit in state $1$ or $0$ and the remaining qubits in the unary state $e_j$ is given by $\Pr[0,e_j] - \Pr[1,e_j] = \frac{1}{4}\left(\frac{1}{\sqrt{n}} + W_j x\right)^2 - \frac{1}{4}\left(\frac{1}{\sqrt{n}} - W_j x\right)^2 = W_j x /\sqrt{n} $. Therefore, for each $j$, we can deduce both the amplitude and the sign of $W_j x$. 

More details about these calculations appear in \cite{klm21}.

Let us provide some remarks about these orthogonal neural networks. First, the quantum circuits have only linear depth and thus one can use a quantum computer to perform inference on the datasets in only linear number of steps. Of course, one can imagine combining such quantum orthogonal layers with other types of quantum layers, in the form of more general parametrized quantum circuits, thus creating more advanced quantum neural network architectures. Second, in \cite{klm21} an efficient classical algorithm for training such orthogonal neural networks is presented that takes the same asymptotic time as training normal dense neural networks, while previously used training algorithms were based on singular value decompositions that have a much worse complexity.
The new training of the orthogonal neural network is done with a gradient descent in the space of the angles of the gates of the quantum circuit and not on the elements of the weight matrix. We denote this training algorithm by QPC, after the quantum pyramid circuit which inspired its creation. Note here that the fact the optimization is performed in the angle landscape, and not on the weight landscape, can produce very different and at times better models, since these landscapes are not the same. This is indeed corroborated by our experimental results as we will discuss below.

\subsection{Hardware demonstration}

\subsubsection{Superconducting quantum computer}

\begin{figure}[!h]
    \centering
    \includegraphics[width=0.35\textwidth]{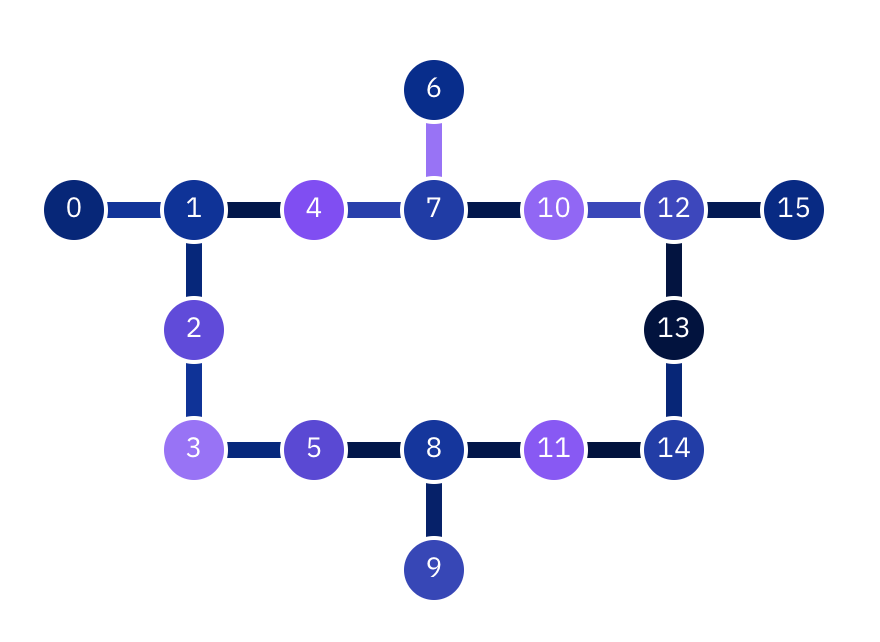}
    \caption{The 16-qubit ibmq\_guadalupe quantum computer.}
    \label{guadalupe}
\end{figure}

The hardware demonstrations were performed on three different superconducting quantum computers provided by IBM, with the majority of the experiments performed on the 16-qubit {\em ibmq\_guadalupe} machine (see Figure \ref{guadalupe}). 

The other two machines that were used were the 7-qubit {\em ibmq\_casablanca} and the 5-qubit {\em ibmq\_bogota}. The quantum hardware was accessed through the IBM Quantum Experience for Business cloud service.

\subsubsection{Quantum software}

The quantum software development was performed using tools from the QC Ware Forge platform, including the quasar language, the data loader and the inner product estimation procedures, as well as the quantum pyramid circuits. The final circuits were translated into qiskit circuits that were then sent to the IBM hardware. 

The datasets were downloaded from the MedMNIST repository \cite{medmnist} and pre-processed by the sklearn implementation of PCA. 

For benchmarking as accurately as possibly against classical fully-connected neural networks, we used the code from \cite{nielsen} for training classical neural networks, and for the quantum-assisted neural networks we adapted the code to use a quantum procedure for the dot product computations. For the orthogonal neural networks, we first designed a new training algorithm for quantum orthogonal neural networks, based on \cite{klm21}, and we also developed code for classical orthogonal neural networks based on singular value decomposition, following \cite{jia2019orthogonal}. 

\subsubsection{Optimizations for the hardware demonstration}

\begin{figure}[h]
\centering
\begin{subfigure}{.5\textwidth}
  \centering
  \includegraphics[width=\linewidth]{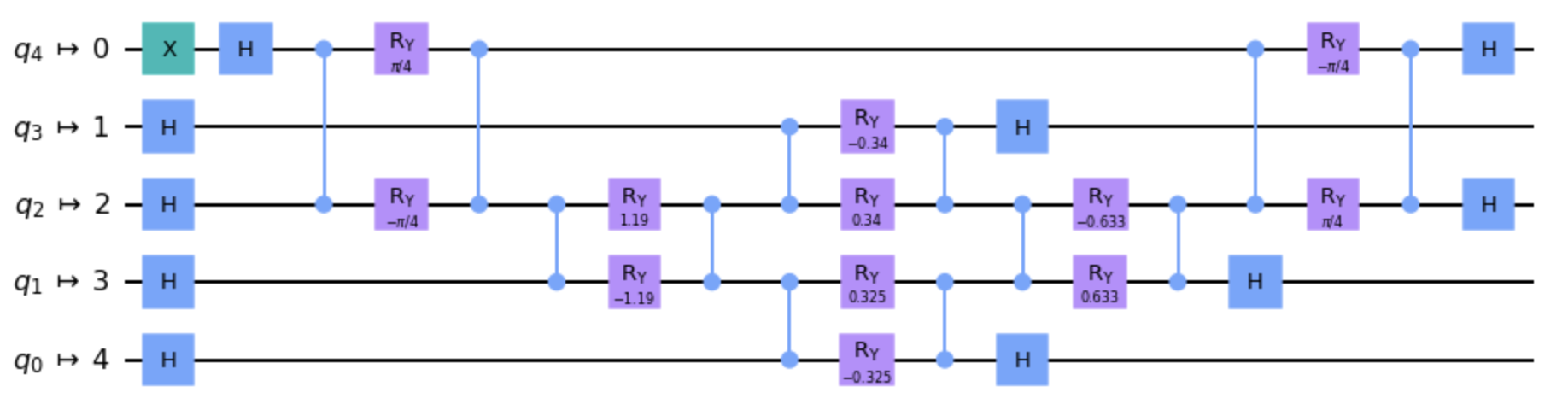}
  \caption{}
  \label{fig:QONNcircuit_rectangular}
\end{subfigure}%
\newline
\begin{subfigure}{.5\textwidth}
  \centering
  \includegraphics[width=\linewidth]{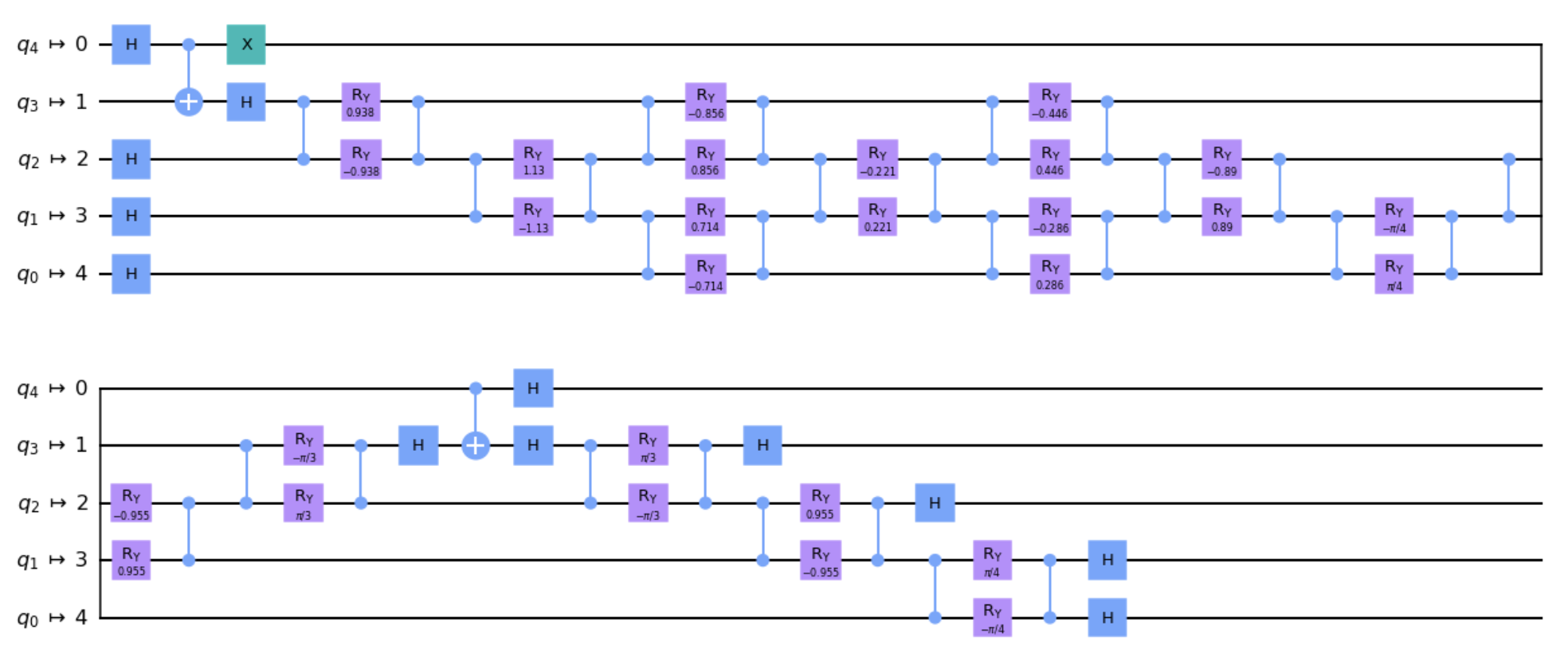}
  \caption{}
\end{subfigure}
\caption{ The optimized qiskit circuits for (a) a [4,2] quantum-assisted NN layer and (b) a [4,2] quantum orthogonal NN layer}
\label{final_circuitIBM}
\end{figure}

Before describing the experimental results, we outline here the optimizations on the circuit design.

First, using a unary encoding for loading the data is very useful in order to mitigate errors that arise from the hardware. In fact all results that correspond to outcomes which are not unary strings can be discarded and this dramatically increases the accuracy of the computation, as we see in the experimental results below.

\begin{figure}[!h]
    \centering
    \includegraphics[width=0.5\textwidth]{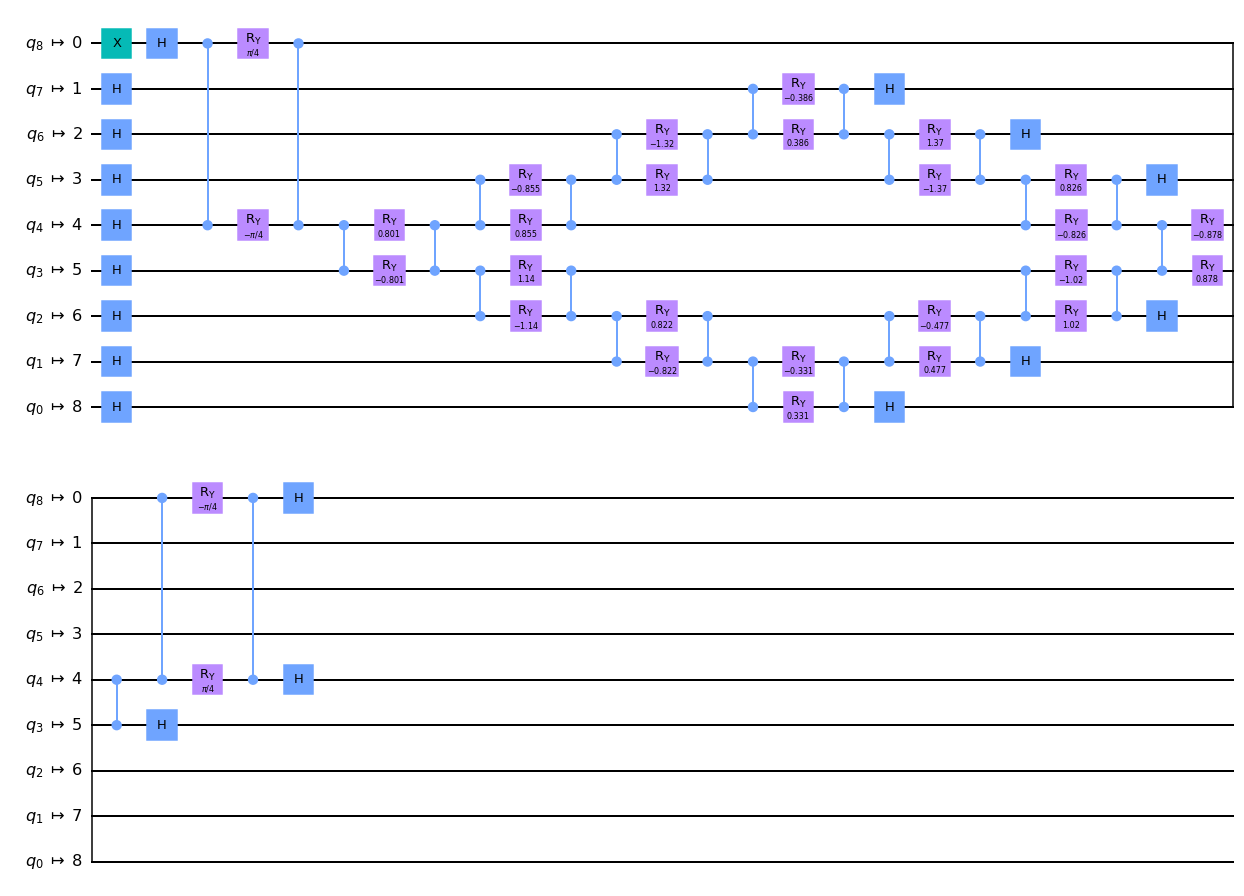}
    \caption{The optimized qiskit circuit for a [8,4] quantum-assisted NN layer}
    \label{qiskit8-4}
\end{figure}

Further optimizations were performed with respect to the layout of the hardware machines and the translation of the RBS gates into native hardware gates. 
We provide in Fig. \ref{qiskit8-4} the final qiskit circuits that were used for a $[8,4]$ layer of the quantum-assisted neural network and of the quantum orthogonal neural network, where a compilation of the original circuits have resulted in a reduction of the number of gates, in particular by removing a number of Hadamard gates that were appearing as pairs. 
The purple boxes correspond to the Ry single qubit gates with a parameter noted within the box.  
We also provide a $[8,2]$ layer of the quantum orthogonal neural network in the Methods.

\subsubsection{Experimental results}

\begin{figure}[!h]
    \centering
    \includegraphics[width=0.85\linewidth]{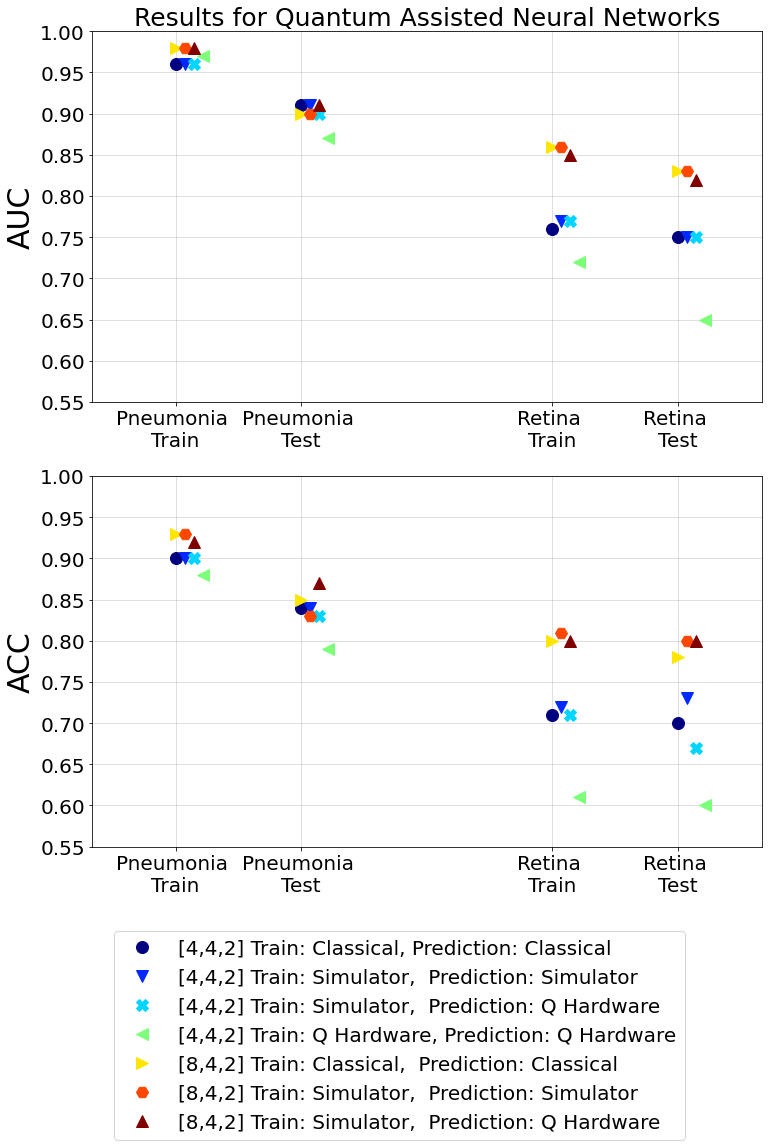}
    \caption{Results of experiments for the Quantum assisted Neural Network.}
    \label{fig:results1}
\end{figure}

\begin{figure}[h]
    \centering
    \includegraphics[width=0.85\linewidth]{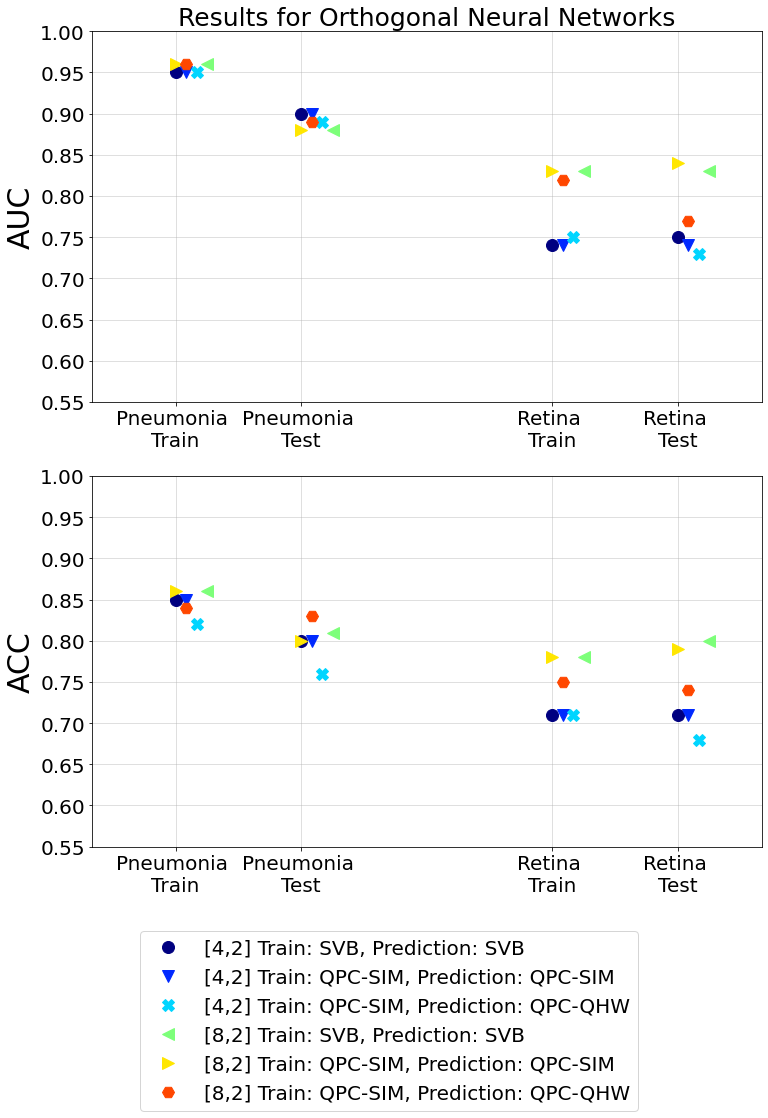}
    \caption{Results of experiments for the Orthogonal Neural Network. QPC stands for Quantum Pyramidal Circuit and is the classical algorithm simulating our quantum circuit. QHW is the quantum circuit on the real quantum hardware. SVB stands for the classical Singular Value Bounded algorithm.}
    \label{fig:results2}
\end{figure}

We tested our methods using a number of different datasets, classification tasks, and architectures that we detail below, and we performed both simulations and hardware experiments. 

In Figures \ref{fig:results1} and \ref{fig:results2} and Table \ref{state}, we show our results, where we provide the AUC (area under curve) and ACC (accuracy) for all different types of neural network experiments, for both the training and test sets, for the Pneumonia and Retina datasets. 



\begin{table*}[]
\begin{tabular}{|cccc|cc|cc|cc|cc|}
\hline
\multicolumn{4}{|c|}{TYPES}              & \multicolumn{2}{c|}{PNEUMONIA AUC} & \multicolumn{2}{c|}{PNEUMONIA ACC} & \multicolumn{2}{c|}{RETINA AUC} & \multicolumn{2}{c|}{RETINA ACC} \\ \hline
method  & layers  & training & inference & TRAIN            & TEST            & TRAIN            & TEST            & TRAIN          & TEST           & TRAIN          & TEST           \\ \hline
qNN     & [4,4,2] & CLA      & CLA       & 0,96        & 0,91       & 0,90        & 0,84       & 0,76      & 0,75      & 0,71      & 0,70      \\
qNN     & [4,4,2] & SIM      & SIM       & 0,96        & 0,91       & 0,90        & 0,84       & 0,77      & 0,75      & 0,72      & 0,73      \\
qNN     & [4,4,2] & SIM      & QHW       & 0,96        & 0,90       & 0,90        & 0,83       & 0,77      & 0,75      & 0,71      & 0,67      \\
qNN     & [4,4,2] & QHW      & QHW       & 0,97             & 0,87            & 0,88             & 0,79            & 0,72           & 0,65           & 0,61           & 0,60           \\
qNN     & [8,4,2] & CLA      & CLA       & 0,98        & 0,90       & 0,93        & 0,85       & 0,86      & 0,83      & 0,80      & 0,78      \\
qNN     & [8,4,2] & SIM      & SIM       & 0,98        & 0,90       & 0,93        & 0,83       & 0,86      & 0,83      & 0,81      & 0,80      \\
qNN     & [8,4,2] & SIM      & QHW       & 0,98             & 0,91            & 0,92             & 0,87            & 0,85           & 0,82           & 0,80           & 0,80           \\ \hline
qOrthNN & [4,2]   & SVB      & SVB       & 0,95        & 0,90       & 0,85       & 0,80       & 0,74      & 0,75      & 0,71      & 0,71      \\
qOrthNN & [4,2]   & QPC-SIM      & QPC-SIM       & 0,95        & 0,90       & 0,85       & 0,80       & 0,74     & 0,74      & 0,71      & 0,71      \\
qOrthNN & [4,2]   & QPC-SIM      & QPC-QHW       & 0,95        & 0,89       & 0,82        & 0,76       & 0,75      & 0,73      & 0,71      & 0,68      \\
qOrthNN & [8,2]   & SVB      & SVB       & 0,96        & 0,88       & 0,86        & 0,81       & 0,83      & 0,83      & 0,78      & 0,80      \\
qOrthNN & [8,2]   & QPC-SIM      & QPC-SIM       & 0,96        & 0,88       & 0,86        & 0,80       & 0,83      & 0,84      & 0,78      & 0,79     \\
qOrthNN & [8,2]   & QPC-SIM      & QPC-QHW       & 0,96        & 0,89       & 0,84        & 0,83       & 0,81      & 0,77      & 0,75      & 0,74      \\ \hline
\end{tabular}
\caption{Results of our experiments on Pneumonia and Retina datasets, reported for the train set and the test set, both using the AUC and ACC metrics. Methods are quantum-assisted neural networks (qNN) or quantum orthogonal neural network (qOrthNN) with different layer architectures. The training and inference are done classically via standard feedforward/backpropagation (CLA), or in a quantum-assisted way both on a quantum simulator (SIM) and on quantum hardware (QHW) for the qNN; and with the Singular-Value Bounded algorithm (SVB), or with a quantum pyramid circuit algorithm both on a quantum simulator (QPC-SIM) or on quantum hardware (QPC-QHW) for the qOrthNN.
The experiments that do not involve quantum hardware have been repeated ten times, the mean values appear on the table and the standard deviation is in most cases $\pm 0.01$ and up to $\pm 0.04$.}
\label{table}
\end{table*}

As general remarks from these results, one can see that for the PneumoniaMNIST dataset, both the AUC and ACC are quite close for all different experiments, showing that the quantum simulations and the quantum hardware experiments reach the performance of the classical neural networks. We also note that the quantum-assisted neural networks achieved somewhat higher accuracies than the orthogonal neural networks. For the RetinaMNIST dataset, the experiments with 8-dimensional data achieve higher AUC and ACC than the ones with 4-dimensional data. The quantum simulations achieve similar performance to the classical one, while for the case where both the training and the inference was performed on a quantum computer, a drop in the performance is recorded, due to the noise present in the hardware and the higher difficulty of the dataset. 

We provide now a more detailed description of the performed experiments.

For the quantum-assisted neural networks, we used two architectures of size $[4,4,2]$ and $[8,4,2]$, meaning the input size is four or eight dimensional respectively, there is one hidden layer of four nodes, we used the sigmoid activation function, and a binary classification task at the end. We also used two small orthogonal neural networks of size $[4,2]$ and $[8,2]$, meaning the input size is four or eight dimensional respectively, there is no hidden layer, we used the sigmoid activation, and a binary classification task at the end. This is due to the fact that the circuit complexity for each layer is still quite high for current quantum machines and better quality qubits would be required in order to scale these architectures up. We performed two classification tasks, one for the PneumoniaMNIST dataset, and one between class 0 and classes $\{1,2,3,4\}$ of the RetinaMNIST dataset. For each task, the training and the inference was performed in combinations of classically, on a quantum simulator, and on quantum hardware. 

The experiments involving classical methods or quantum simulators have been repeated ten times each to extract mean values and error bars for the AUC and ACC quantities. The mean values appear in Table \ref{table} and the errors are bounded by $\pm 0.01 $ for most cases and up to $\pm 0.04$, which reflects the randomness in initializing the weights and randomness in the quantum estimation procedures. For the hardware experiments, we performed the experiments once per different type of neural network, of layer architecture, of training method, of inference method and on the training and test sets of the RetinaMNIST and PneumoniaMNIST datasets. The hardware experiments took between 45 minutes to several hours. The longest one, training the $[4,4,2]$ quantum-assisted neural network, took more than 10 hours, with the majority of this time not on the actual quantum hardware but on handling a really large number of jobs within the quantum cloud service. The training and inference with classical methods or the quantum simulator took a few seconds to complete.

In order to benchmark more precisely the performance of the quantum hardware for the different types of quantum circuits we used, we provide in Methods an analysis, where we plot the simulated versus the experimental value of the output of the quantum circuits we used in the quantum-assisted neural networks over the entire test sets. This allows us to  see how the quantum hardware behaves on real data. For the 5 qubit experiments the hardware results are very close to the exepcted ones, while for the 9-qubit experiments we see that with some non-trivial probability the hardware execution diverges from the simulation. 

We note also that the  simulations show that the models trained through quantum methods are robust and the differences in AUC and ACC between different runs are small, and comparable to the classical models. On the other hand, the behaviour of the quantum hardware can be quite unstable over time, where repeating an experiment in different days or weeks can provide quite different results, not due to the randomness in the training or inference (these differences can be seen from the simulation results to be very small, namely about $\pm 0.01$), but due to the levels of noise of the hardware that is better or worse calibrated at different time periods or due to which specific subset of qubits one uses within the same quantum computer, or which quantum hardware machine one uses.

\paragraph{Simulation results}

\begin{figure}[!h]
    \centering
    \includegraphics[width=0.5\textwidth]{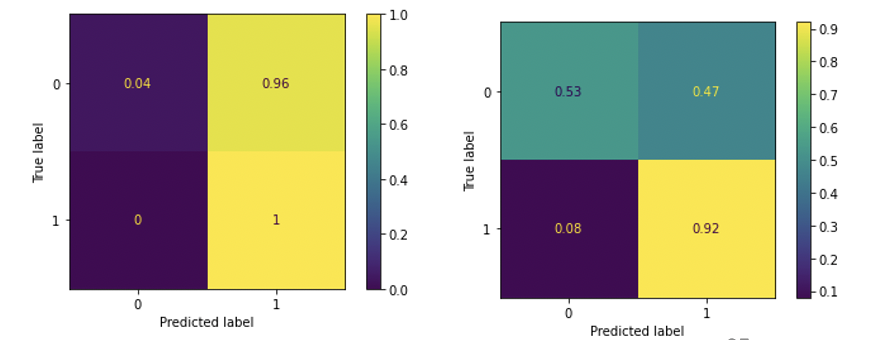}
    \caption{Results of training of a [32,16,2] orthogonal neural network on the RetinaMNIST (classes 0 vs. \{1,2,3,4\}) with training set of size 1080 (486-594) and test size 400 (174-226) with the SVB-based algorithm (left side: ACC (test set) = 58.25\%) and the simulated quantum pyramid circuit algorithm (right side: ACC (test set) = 75.25
}
    \label{diffmodel}
\end{figure}

Looking at the quantum simulation results, the AUC and ACC of the quantum-assisted neural networks match those of the corresponding classical ones. This is to be expected, since the quantum circuits are assisting in the training but are not changing the classical architectures. The main difference stems from quantum procedures which estimate and not compute exactly quantities such as inner products between data points and weights. This way, the quantum-assisted neural-nets are closer to classical FNNs where noise has been injected artificially during training and inference. Such noise injection can actually be beneficial at times, especially in the presence of small training sets, since they make the models more robust and potentially generalize better.
For the quantum orthogonal neural networks, the performance varied more compared to the performance of same-size classical orthogonal neural networks based on Singular Value Decomposition (SVB), and this difference can be explained by the training methods which are completely different. The new quantum-inspired way of performing classical training, using the parameters of the gates of the quantum circuit instead of the elements of the weight matrices, as described in \cite{klm21}, allows for a training in time $O(n^2)$ instead of the previously known $O(n^3)$ \cite{jia2019orthogonal}. Moreover, the models produced are quite different due to the gradient optimization on the landscape of the circuit parameters and thus can be a powerful source of more accurate models. We provide a specific example of a substantially different training in Figure \ref{diffmodel}, where we see substantial difference in the ACC and confusion matrices, where the quantum-inspired way of training achieves $75\%$ accuracy and the SVB-based one does not train and outputs practically always 1. 

\begin{figure}[!h]
    \centering
    \includegraphics[width=0.5\textwidth]{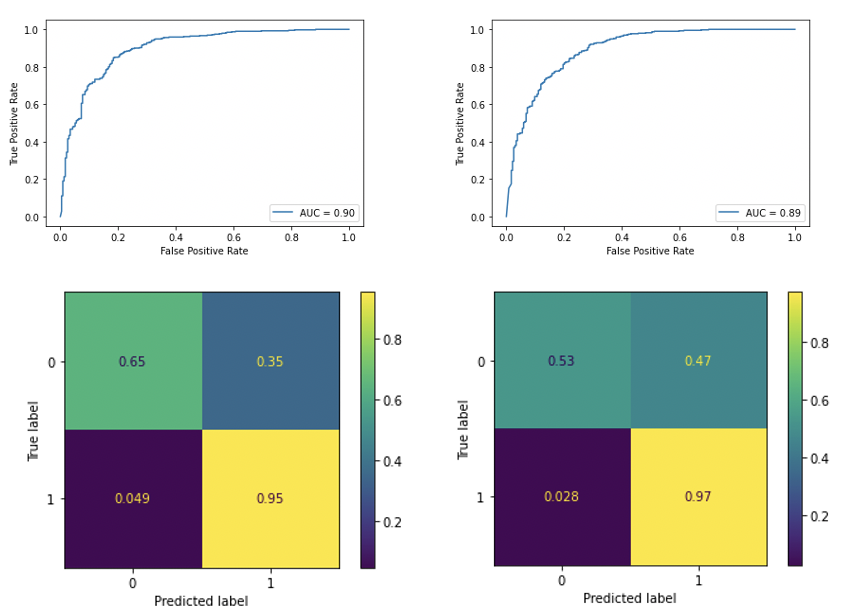}
    \caption{Training of a [784,64,2] quantum-assisted neural network on the PneumoniaMNIST with training set of size 4708 (1214 – 3494) and test size of 624 (234-390) with classical training (left side: AUC (test set) = 0.90, ACC (test set) = 83.65\%) and the simulated quantum-assisted training algorithm (right side: AUC (test set) = 0.89, ACC (test set) = 80.77\%).
}
    \label{bigsim}
\end{figure}

Thanks to tailor-made quantum simulators for the specific circuits we use, we were able to perform larger-scale simulations which provide strong evidence of the scalability of our methods. In Figure \ref{bigsim} we provide an example of simulation results for quantum-assisted neural networks on the 784 dimensional images that match the accuracy of the classical NNs. 

Last, in Methods we provide simulations of the training of the quantum orthogonal neural networks for different layer sizes to show in practice the asymptotic running time of $O(n^2)$ of the quantum pyramid circuit training algorithm. We also provide more details about the number of steps that the quantum-assisted neural networks take in comparison to the classical training.

\paragraph{Hardware results}

Looking at the results of the hardware demonstration, it is clear that current hardware is not ready to perform medical image classification in a way that is competitive to classical neural networks, since the hardware does not have sufficient number and quality of qubits. Possibly better algorithms and heuristics are needed to train better and faster models. 
Nevertheless, the experiments showed promising results with 5-qubit and to a lesser extent with 9-qubit circuits, providing small-scale confirmation of the methods, while the larger simulations we performed provide more evidence about the scalability and future performance. 

One can see the overall results of the simulations and hardware experiments in Table \ref{table}.

Overall, classically-trained architectures and quantum inference matches the classical and quantum simulator performance, while when we performed both training and inference on a quantum computer, the $[4,4,2]$ neural network managed to train with a small drop in the accuracy, while the $[8,4,2]$ neural network did not train, outputting always the same classification value for the entire dataset. 

For the orthogonal neural networks, the results of the hardware demonstration match the quantum simulation ones (with a small drop only for the RetinaMNIST and the 9-qubit experiments), showing that overall the hardware has sufficient quality to perform the necessary quantum circuits. Note that in this case, the training can always be performed on a classical computer with the optimal training algorithm developed in \cite{klm21}. 

We provide some further results, including AUC, ACC and confusion matrices for a number of the experiments in the Methods.

\section{Methods}


\subsection{Additional qiskit circuits}

We provide here the circuits for the $[8,4]$ layer of the quantum-assisted neural network and a $[8,2]$ layer of the quantum orthogonal neural network. We have seen above how these circuits scale for larger dimensions. 

\begin{figure}[!h]
    \centering
    \includegraphics[width=0.5\textwidth]{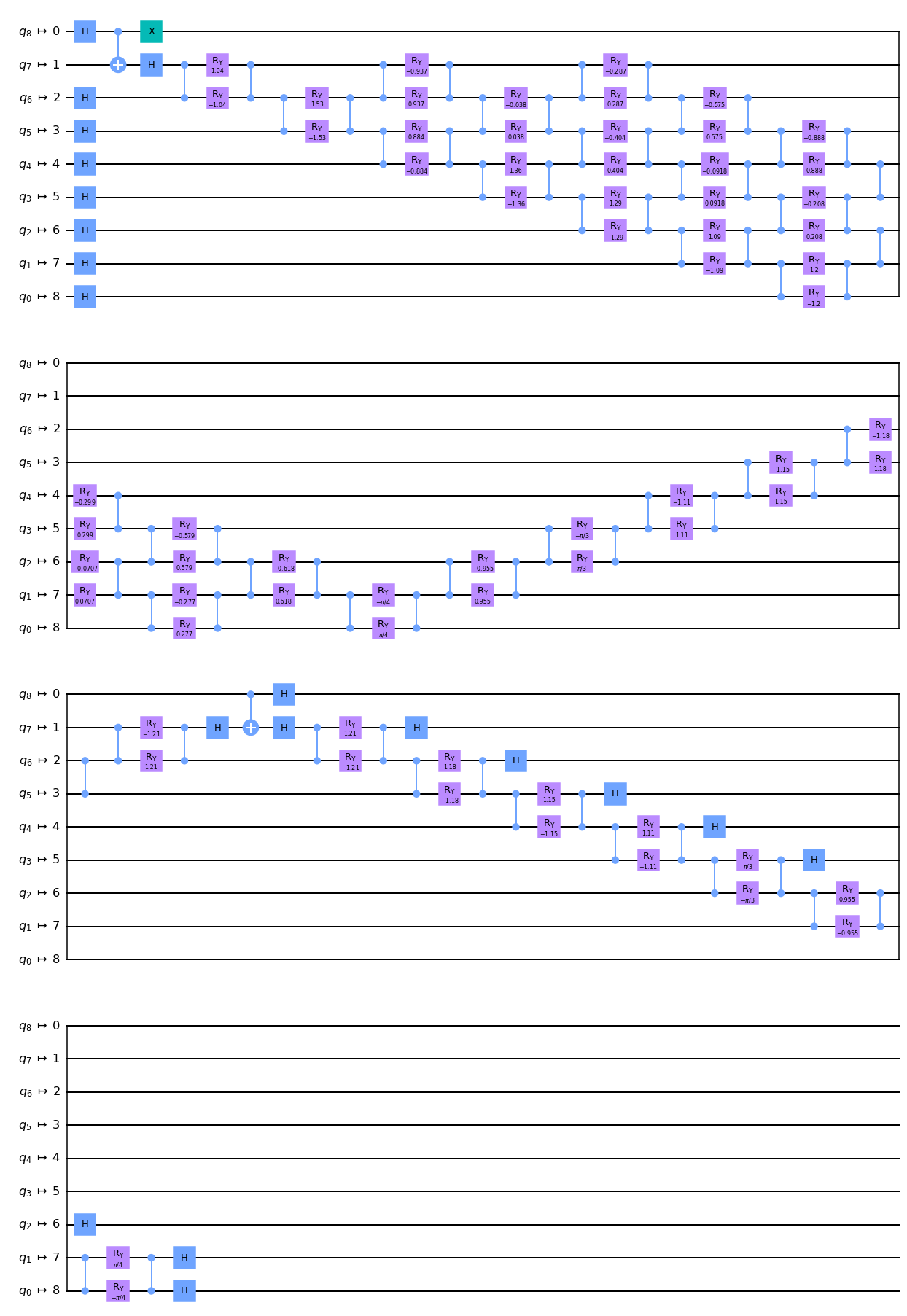}
    \caption{The optimized qiskit circuit for a [8,2] quantum orthogonal NN layer}
    \label{qiskit8-2}
\end{figure}

\subsection{Additional simulation results}

\begin{figure}[!h]
    \centering
    \includegraphics[width=0.5\textwidth]{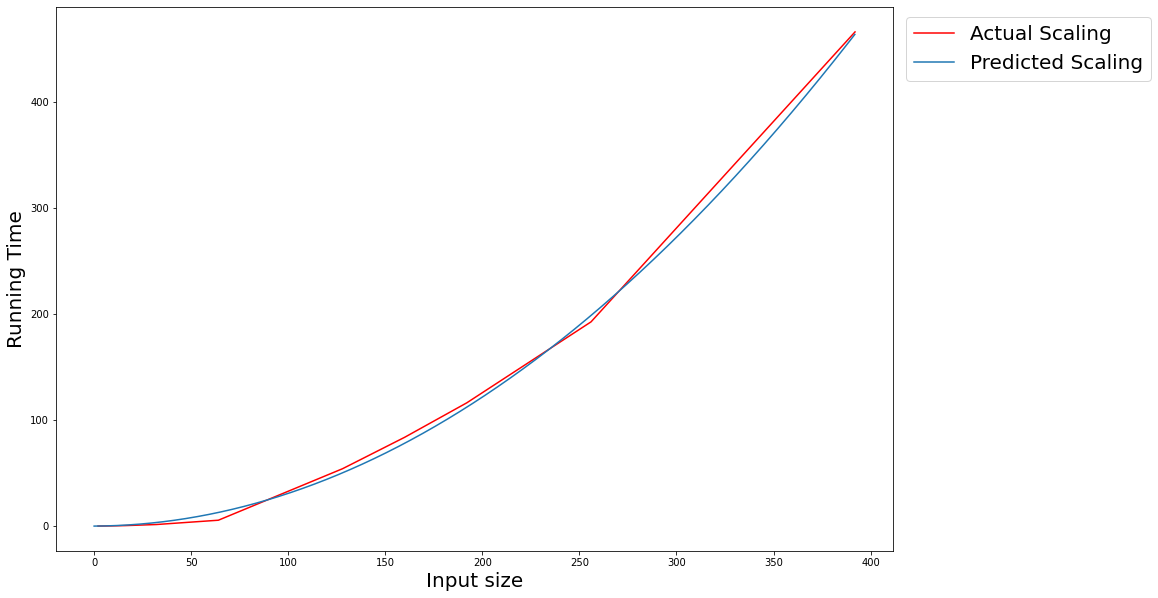}
    \caption{The actual scaling of the running time for growing dimension $n$ of training a $[n,n,2]$ orthogonal neural network and the predicted scaling 0.006($0.5 n^2 + 1.5n - 3$).}
    \label{time}
\end{figure}

As was proved in \cite{klm21}, the scaling of the running time of the training of the orthogonal neural networks based on the quantum pyramid circuit is linear with respect to the number of parameters. This is corroborated by our results here as shown in Figure \ref{time}. In particular, we trained $[n,n,2]$ quantum orthogonal neural networks, for different values of $n \in [2,392]$ and saw that the running time grows indeed as the number of parameters which is  $0.5 n^2 + 1.5n - 3$. This is asymptotically better than the previously known training algorithms for orthogonal neural networks that run in time $O(n^3)$. 

\begin{figure}[!h]
    \centering
    \includegraphics[width=0.5\textwidth]{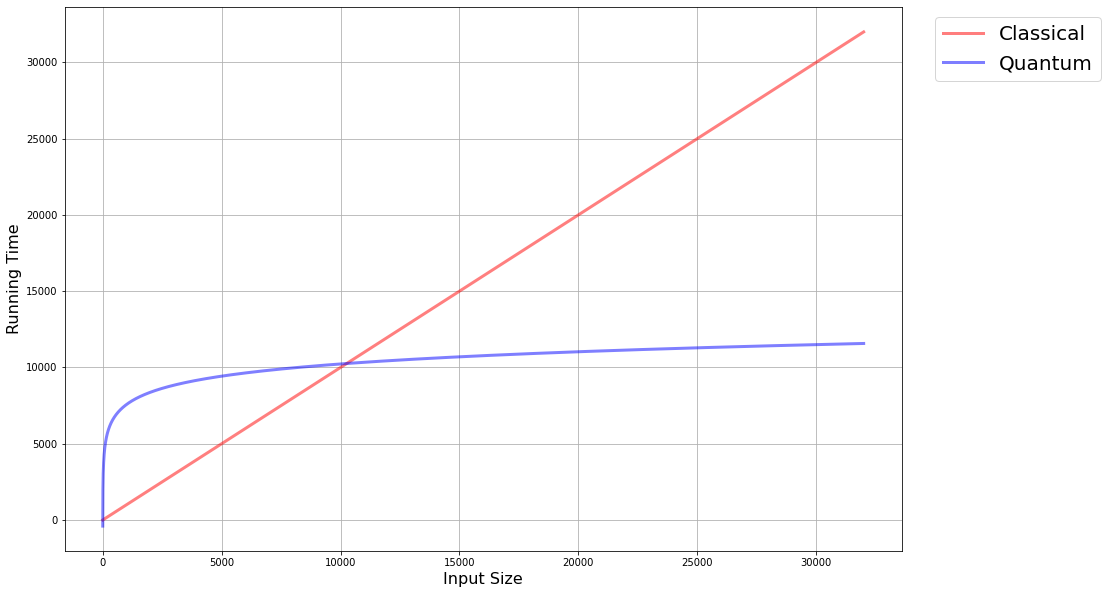}
    \caption{The scaling of quantum steps for the inner 
product estimation between two n-dim vectors  
(blue: $400(2\log n-1)$)  versus for the classical inner 
product computation (red: $n$). The crossover point is for $100 \times 100$ pixel images.}
    \label{time_qnn}
\end{figure}

Note as well that the running time of the quantum-assisted neural networks can be analysed theoretically as in \cite{NearestCentroid2021}, where the main difference with classical fully-connected neural networks is on the computation of the inner product between vectors. While classically an inner-product computation between two $n$-dimensional vectors takes $n$ steps, a quantum circuit for estimating the inner product that uses the parallel loaders from Fig. \ref{loaders} has depth of only $2\log(n)-1$. These shallow quantum circuits need to be repeated a number of times (shots) in order to get a accurate estimation of the inner product, and in theory to get an $\epsilon$-accurate result the total number of steps of the quantum circuits will be $O(\log n/\epsilon^2)$. We find that repeating the quantum circuits 400 times (independent of the dimension $n$) suffices to get the desired accuracies. In Fig. \ref{time_qnn} we compare the scaling of the steps of the quantum versus classical computation and see that for images of size $100 \times 100$ the quantum steps start becoming fewer. A smaller number of steps does not imply immediately a faster running time, since one needs to assume that in the future a single quantum processor can apply gates on different qubits in parallel (see e.g. \cite{parallelGates}) and also take into account the time to apply one quantum or one classical step. 

\subsection{Additional hardware results}

\begin{figure}[!h]
    \centering
    \includegraphics[width=0.5\textwidth]{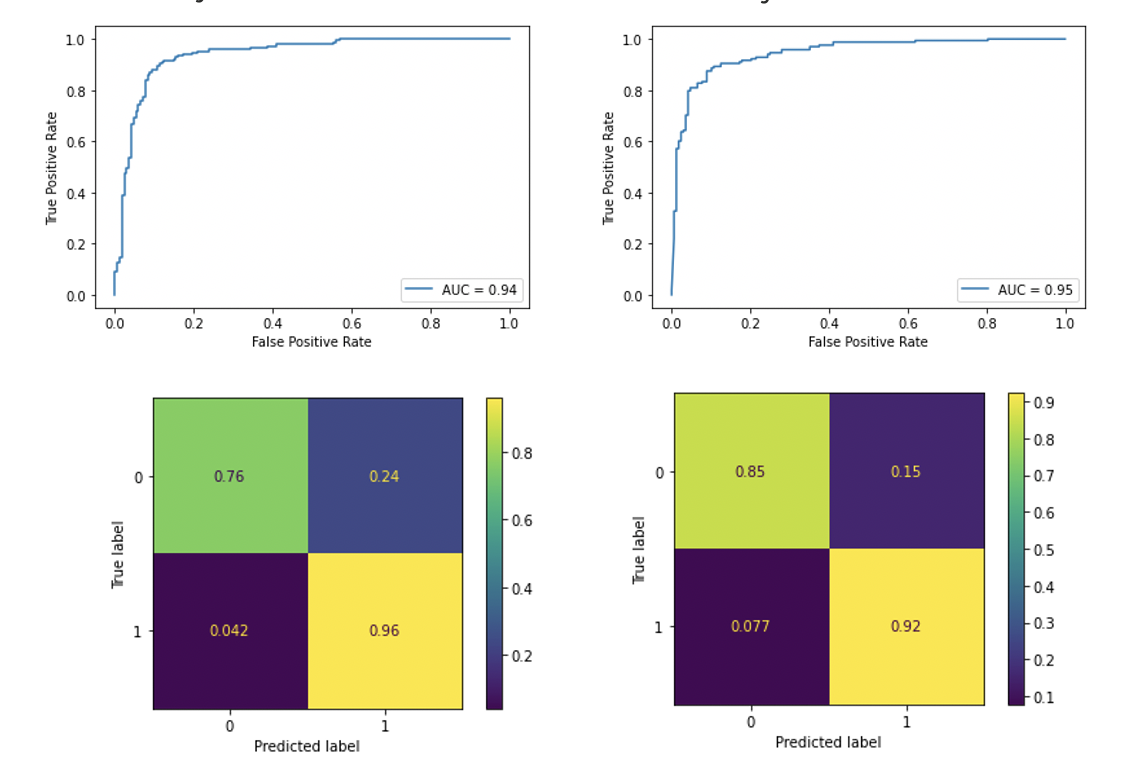}
    \caption{Training of a [8,4,2] quantum-assisted neural network on the PneumoniaMNIST with training set of size 2428 (1214 – 1214)  and test size 336 (168-168) with training on a quantum simulator and inference on a quantum hardware (left: AUC (test set) = 0.94, ACC = (test set) 86.01\%), and on the quantum simulator(right: AUC = (test set) 0.95, ACC (test set) = 88.39\%).
}
    \label{hwqNNPneumonia}
\end{figure}

\begin{figure}[!h]
    \centering
    \includegraphics[width=0.5\textwidth]{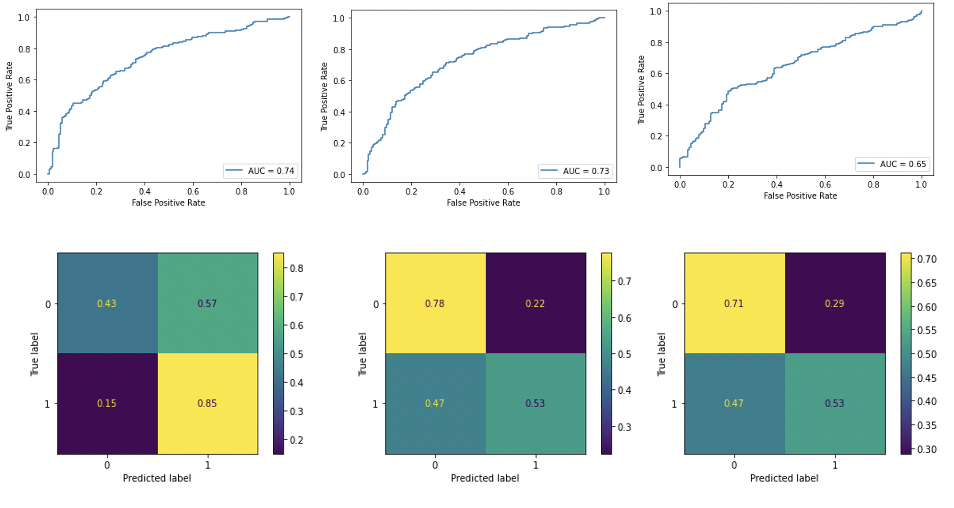}
    \caption{Training of a [4,4,2] quantum-assisted neural network on the RetinaMNIST with training set of size 100 (50-50) and test size 400 (174-226) with both training and testing on a quantum simulator (left: AUC (test set) = 0.74, ACC (test set) = 66.5\%), the training on a quantum simulator and the inference on the quantum hardware (middle: AUC (test set) = 0.73, ACC (test set) = 63.75\%), and last with both the training and testing on the quantum hardware (right: AUC (test set) = 0.65, ACC (test set) = 60\%).
}
    \label{hwqNNRetina}
\end{figure}

\begin{figure}[!h]
    \centering
    \includegraphics[width=0.5\textwidth]{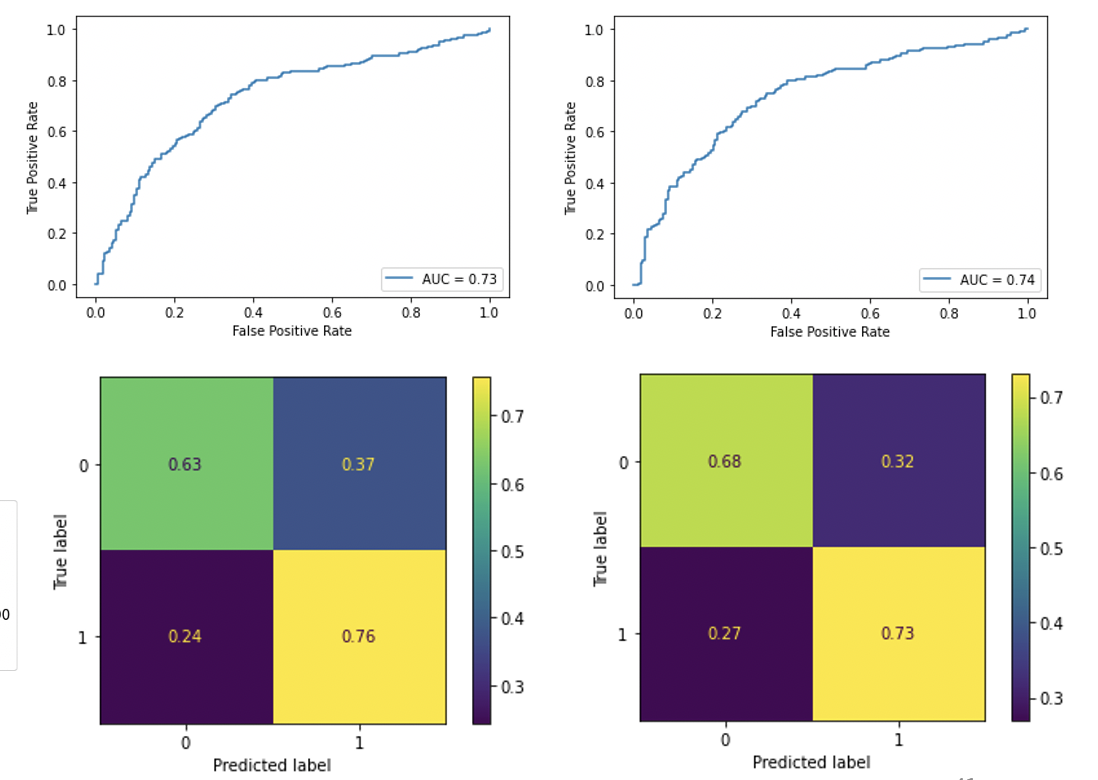}
    \caption{Training of a [4,2] quantum orthogonal neural network on the RetinaMNIST with training set of size 1080 (486-594) and test size 400 (174-226) with testing on a quantum simulator (left: AUC (test set) = 0.73, ACC (test set) = 70\%), and testing on the quantum hardware (right: AUC (test set) = 0.74, ACC (test set) = 70.75\%).
}
    \label{hwqOrthNNRetina}
\end{figure}

In Fig. \ref{hwqNNPneumonia},\ref{hwqNNRetina}, and \ref{hwqOrthNNRetina} we provide the AUC curves, ACC and confusion matrices for some more of the experiments we performed, including both the PneumoniaMNIST and RetinaMNIST datasets, both quantum-assisted and orthogonal neural networks, and for training and inference on simulators or quantum hardware. 

We also present here an analysis for the quantum circuits we used in order to perform the quantum-assisted neural networks, namely the quantum inner product estimation circuits, which are useful for other applications beyond training neural networks. 

In particular, we present inference results from the 624 data points in the test set of the PneumoniaMNIST and the 400 data points in the test set of the RetinaMNIST, and also both with the [4,4,2] and [8,4,2] quantum-assisted neural networks, and we check the estimated value of the first node in the final layer of the neural network (which corresponds to the first class of the binary classification task) on the simulator versus on the real hardware. The weights we used were found by performing the training on the quantum simulator. This allows us to perform the same quantum circuits a large number of times on different inputs and get an estimate of how well the quantum hardware can perform this particular application task.

\begin{figure}[!h]
    \centering
    \includegraphics[width=0.4\textwidth]{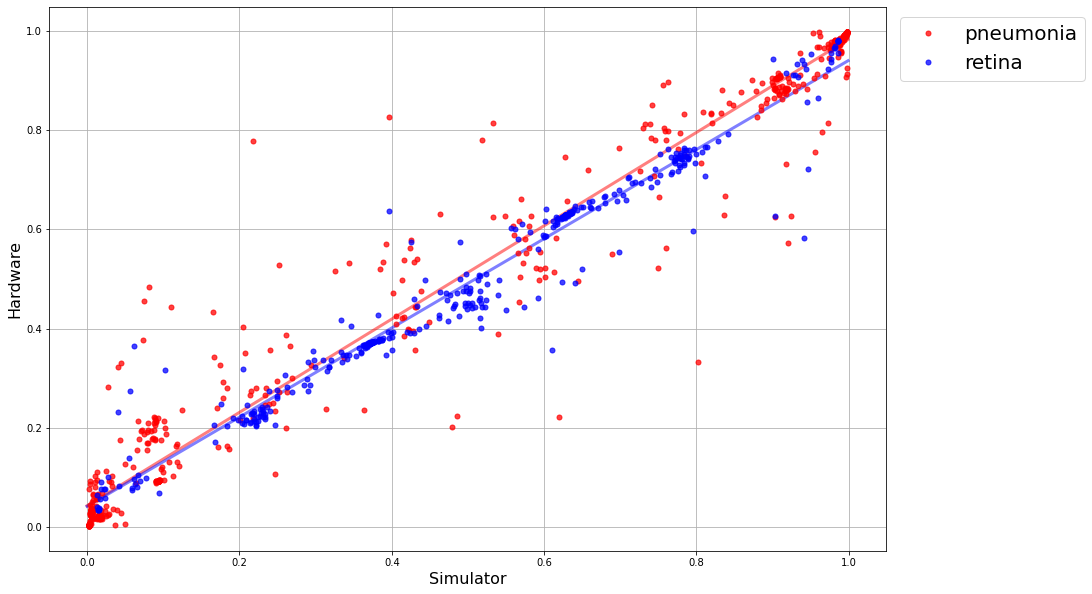}
    \caption{Experimental versus simulated value of the first node of the final layer of the [4,4,2] qNN. Pneumonia: slope=$0.94 \pm 0.09$, intercept=$0.04\pm 0.06$; Retina: slope=$0.90\pm 0.20$, intercept=$0.04 \pm 0.14$
}
    \label{4-hw}
\end{figure}

\begin{figure}[!h]
    \centering
    \includegraphics[width=0.4\textwidth]{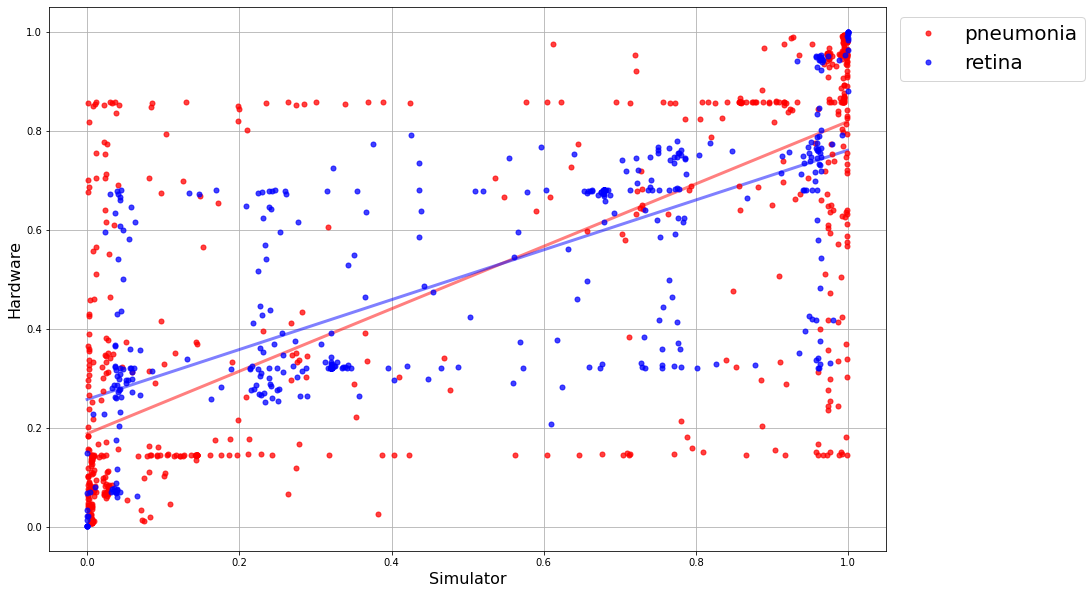}
    \caption{Experimental versus simulated value of the first node of the final layer of the [8,4,2] qNN. Pneumonia: slope=$0.63 \pm 0.09$, intercept=$0.19\pm 0.06$; Retina: slope=$0.50\pm 0.15$, intercept=$0.26 \pm 0.10$}
    \label{8-hw}
\end{figure}

In Figure \ref{4-hw}, we see that for the $[4,4,2]$ architecture the results of the hardware executions are quite close to the expected ones from the simulation for both data sets. In Figure \ref{8-hw}, we see that for the $[8,4,2]$ architecture, while most of the hardware results agree with the simulations, there is a non-trivial fraction of the points where an error occurs in the quantum circuit that causes the results between simulation and hardware execution to diverge. This does not necessarily translate to a drop in accuracy, this indeed does not happen as we see in Table \ref{table}, since some points that were mis-classified by the simulator can now be classified correctly on the quantum hardware due to possible errors in the quantum circuits. It is quite clear that the larger circuits we used are pushing the boundaries of what the current quantum machines can perform reliably and one would need better and more robust hardware to increase confidence on how quantum machine learning techniques can help in image classification tasks.

\section{Discussion}

We provided here an initial study of quantum neural network techniques for medical image classification.
We initiated the use of the MedMNIST suite of datasets as a clear, standardized and lightweight benchmark for classification via quantum techniques and we showed via simulations and hardware demonstrations how quantum-assisted neural networks and quantum orthogonal neural networks can be powerful tools for medical image classification. Our current methods have been designed in a way to retain a clear connection to classical methods, in order to be able to study theoretically their scalability, performance and training time. They can of course be combined with other types of parametrized quantum circuits that explore larger Hilbert spaces, though if and how such techniques can enhance the training of models for classical data is still an open question.
The simulations and hardware experiments we performed were extensive with more than a billion shots in total, showing the power and limitations of current hardware.

Regarding the potential advantages of quantum neural network techniques:
as far as speed is concerned, with the coming of faster quantum machines that will also have the ability to perform parallel gates on qubits, we can expect faster quantum training and inference via the log-depth quantum inner product estimation circuits and more importantly with more advanced quantum procedures for linear-algebra. However, highly specialized classical hardware, e.g. GPUs, are extremely efficient in matrix-vector multiplications, so it is very likely that more powerful quantum techniques will have to be combined with the simpler techniques described here. We note also that the study of quantum computing can also lead to novel, faster classical training methods, as is the case for orthogonal neural networks \cite{klm21}.

With respect to the accuracy of the quantum models, we see the potential of quantum neural networks to provide different and at times better models, by having access to different optimization landscapes that could avoid effects like barren plateaus due to quantum properties like, for example, orthogonality. 
Note also, that there is a possibility for the quantum-assisted neural networks to actually provide different models via gradient optimization on the landscape of the circuit parameters and not directly on the elements of the weight matrices. We performed such gradient computations for the more complex orthogonal neural network case, and the techniques can be readily transferred to this easier case. Whether such different quantum models are advantageous will depend on the specifics of the use case and further experimentation is certainly needed.

Last, the application of neural networks is ubiquitous in almost all data domains, not only in image classification, and, thus, we expect the quantum methods developed and tested in this work to have much broader impact both in life sciences and beyond. Such methods will of course continue to be refined as the hardware evolves.


\section{Data Availability}
The data supporting the findings is available from the corresponding author upon reasonable request.

\section{Author Contribution}

NM developed code for the quantum neural networks, compiled the quantum circuits, and performed the corresponding experiments and data analysis. JL developed code for the orthogonal neural networks and performed data analysis. YL and MS formulated the use case and relevant experiments to demonstrate usefulness of this work. SK has developed code for the different types of data loaders. AP has contributed in the theoretical analysis of the quantum neural networks. IK conceived the project and contributed in all aspects.

\section{Acknowledgements}
This work is a collaboration between QC Ware and Roche. We acknowledge the use of IBM Quantum services for this work. The views expressed are those of the authors, and do not reflect the official policy or position of IBM or the IBM Quantum team.
The following members of the Roche pRED Quantum Computing Taskforce also contributed to this work: Marielle van de Pol, Agnes Meyder, Detlef Wolf, Stanislaw Adaszewski.

\vspace{0.5cm}
\section{Competing Interests}
\vspace{-0.5cm}
The authors declare no competing interests.

\bibliographystyle{plain}
\bibliography{references}

\begin{thebibliography}{10}

\bibitem{retina}
Deepdr diabetic retinopathy image dataset (deepdrid), "the 2nd diabetic
  retinopathy – grading and image quality estimation challenge,".
\newblock {\em https://isbi.deepdr.org/data.html}, 2020.

\bibitem{abbas2021power}
Amira Abbas, David Sutter, Christa Zoufal, Aur{\'e}lien Lucchi, Alessio
  Figalli, and Stefan Woerner.
\newblock The power of quantum neural networks.
\newblock {\em Nature Computational Science}, 1(6):403--409, 2021.

\bibitem{QNN2020}
J~Allcock, CY~Hsieh, I~Kerenidis, and S~Zhang.
\newblock Quantum algorithms for feedforward neural networks.
\newblock {\em ACM Transactions on Quantum Computing 1 (1), 1-24}, 2020.

\bibitem{beer2020training}
Kerstin Beer, Dmytro Bondarenko, Terry Farrelly, Tobias~J Osborne, Robert
  Salzmann, Daniel Scheiermann, and Ramona Wolf.
\newblock Training deep quantum neural networks.
\newblock {\em Nature communications}, 11(1):1--6, 2020.

\bibitem{Benjamens}
Stan Benjamens, Pranavsingh Dhunnoo, and Bertalan Mesko.
\newblock The state of artificial intelligence-based fda-approved medical
  devices and algorithms: an online database.
\newblock {\em npj Digital Medicine volume 3, Article number: 118}, 2020.

\bibitem{bharti2021noisy}
Kishor Bharti, Alba Cervera-Lierta, Thi~Ha Kyaw, Tobias Haug, Sumner
  Alperin-Lea, Abhinav Anand, Matthias Degroote, Hermanni Heimonen, Jakob~S
  Kottmann, Tim Menke, et~al.
\newblock Noisy intermediate-scale quantum (nisq) algorithms.
\newblock {\em arXiv preprint arXiv:2101.08448}, 2021.

\bibitem{Polyadic2020}
William Cappelletti, Rebecca Erbanni, and Joaquín Keller.
\newblock Polyadic quantum classifier.
\newblock {\em arXiv:2007.14044}, 2020.

\bibitem{VQCA2020}
M~Cerezo, A~Sone, L~Cincio, and P~Coles.
\newblock Barren plateau issues for variational quantum-classical algorithms.
\newblock {\em Bulletin of the American Physical Society 65}, 2020.

\bibitem{QCNNplateaus2020}
M.~Cerezo, Akira Sone, Tyler Volkoff, Lukasz Cincio, and Patrick~J. Coles.
\newblock Cost-function-dependent barren plateaus in shallow quantum neural
  networks.
\newblock {\em arXiv:2001.00550}, 2020.

\bibitem{xiaodi_neurips}
Shouvanik Chakrabarti, Huang Yiming, Tongyang Li, Soheil Feizi, and Xiaodi Wu.
\newblock Quantum wasserstein generative adversarial networks.
\newblock In {\em Advances in Neural Information Processing Systems},
  volume~32. Curran Associates, Inc., 2019.

\bibitem{CNN2019}
Iris Cong, Soonwon Choi, and Mikhail~D. Lukin.
\newblock Quantum convolutional neural networks.
\newblock {\em Nature Physics 15}, 2019.

\bibitem{cong2019quantum}
Iris Cong, Soonwon Choi, and Mikhail~D Lukin.
\newblock Quantum convolutional neural networks.
\newblock {\em Nature Physics}, 15(12):1273--1278, 2019.

\bibitem{coyle2020born}
Brian Coyle, Daniel Mills, Vincent Danos, and Elham Kashefi.
\newblock The born supremacy: quantum advantage and training of an ising born
  machine.
\newblock {\em npj Quantum Information}, 6(1):1--11, 2020.

\bibitem{QNN2018}
Edward Farhi and Hartmut Neven.
\newblock Classification with quantum neural networks on near term processors.
\newblock {\em arXiv:1802.06002}, 2018.

\bibitem{Image2020}
Hector~Ivan Garcıa-Hernandez, Raymundo Torres-Ruiz, and Guo-Hua Sun.
\newblock Image classification via quantum machine learning.
\newblock {\em arXiv:2011.02831}, 2020.

\bibitem{Hierarchical2018}
Edward Grant, Marcello Benedetti, Shuxiang Cao, Andrew Hallam, Joshua Lockhart,
  Vid Stojevic, Andrew~G. Green, and Simone Severini.
\newblock Hierarchical quantum classifiers.
\newblock {\em npj Quantum Information 4, arXiv:1804.03680}, 2018.

\bibitem{parallelGates}
Nikodem Grzesiak, Reinhold Blümel, Kenneth Wright, Kristin~M. Beck, Neal~C.
  Pisenti, Ming Li, Vandiver Chaplin, Jason~M. Amini, Shantanu Debnath, Jwo-Sy
  Chen, and Yunseong Nam.
\newblock Efficient arbitrary simultaneously entangling gates on a trapped-ion
  quantum computer.
\newblock {\em Nat Commun, 11}, 2020.

\bibitem{Supervised2018}
Vojtech Havlicek, Antonio~D. Córcoles, Kristan Temme, Aram~W. Harrow, Abhinav
  Kandala, Jerry~M. Chow, and Jay~M. Gambetta.
\newblock Supervised learning with quantum enhanced feature spaces.
\newblock {\em Nature volume 567, arXiv:1804.11326}, 2018.

\bibitem{jia2019orthogonal}
Kui Jia, Shuai Li, Yuxin Wen, Tongliang Liu, and Dacheng Tao.
\newblock Orthogonal deep neural networks.
\newblock {\em IEEE transactions on pattern analysis and machine intelligence},
  2019.

\bibitem{NearestCentroid2021}
S~Johri, S~Debnath, A~Mocherla, A~Singh, A~Prakash, J~Kim, and I~Kerenidis.
\newblock Nearest centroid classification on a trapped ion quantum computer.
\newblock {\em npj Quantum Information (to appear), arXiv:2012.04145}, 2021.

\bibitem{QCNN2019}
I~Kerenidis, J~Landman, and A~Prakash.
\newblock Quantum algorithms for deep convolutional neural networks.
\newblock {\em EIGHTH INTERNATIONAL CONFERENCE ON LEARNING REPRESENTATIONS
  ICLR}, 2019.

\bibitem{klm21}
Iordanis Kerenidis, Jonas Landman, and Natansh Mathur.
\newblock Classical and quantum algorithms for orthogonal neural networks.
\newblock {\em arXiv:2106.07198}, 2021.

\bibitem{pneumonia}
Daniel~S. Kermany, Michael Goldbaum, and et~al.
\newblock Identifying medical diagnoses and treatable diseases by image-based
  deep learning.
\newblock {\em Cell, vol. 172, no. 5, pp. 1122 – 1131.e9,}, 2018.

\bibitem{Kiani20}
Bobak~Toussi Kiani, Agnes Villanyi, and Seth Lloyd.
\newblock Quantum medical imaging algorithms.
\newblock {\em arXiv:2004.02036}, 2020.

\bibitem{Dressed2020}
Saurabh Kumar, Siddharth Dangwal, and Debanjan Bhowmik.
\newblock Supervised learning using a dressed quantum network with "super
  compressed encoding": Algorithm and quantum-hardware-based implementation.
\newblock {\em arXiv:2007.10242}, 2020.

\bibitem{Landman2022quantummethods}
Jonas Landman, Natansh Mathur, Yun~Yvonna Li, Martin Strahm, Skander Kazdaghli,
  Anupam Prakash, and Iordanis Kerenidis.
\newblock Quantum {M}ethods for {N}eural {N}etworks and {A}pplication to
  {M}edical {I}mage {C}lassification.
\newblock {\em {Quantum}}, 6:881, December 2022.

\bibitem{mitarai2018quantum}
Kosuke Mitarai, Makoto Negoro, Masahiro Kitagawa, and Keisuke Fujii.
\newblock Quantum circuit learning.
\newblock {\em Physical Review A}, 98(3):032309, 2018.

\bibitem{Semisupervised2020}
Kouhei Nakaji and Naoki Yamamoto.
\newblock Quantum semi-supervised generative adversarial network for enhanced
  data classification.
\newblock {\em arXiv:2010.13727}, 2020.

\bibitem{nielsen}
Michael~A. Nielsen.
\newblock Neural networks and deep learning.
\newblock {\em Determination Press}, 2015.

\bibitem{nosarzewski2018deep}
Benjamin Nosarzewski.
\newblock Deep orthogonal neural networks.
\newblock 2018.

\bibitem{unary2019}
Sergi Ramos-Calderer, Adrián Pérez-Salinas, Diego García-Martín, Carlos
  Bravo-Prieto, Jorge Cortada, Jordi Planagumà, and José~I. Latorre.
\newblock Quantum unary approach to option pricing.
\newblock {\em arXiv:1912.01618}, 2019.

\bibitem{Train2020}
K~Sharma, M~Cerezo, L~Cincio, and PJ~Coles.
\newblock Trainability of dissipative perceptron-based quantum neural networks.
\newblock {\em arXiv preprint arXiv:2005.12458}, 2020.

\bibitem{wang2020orthogonal}
Jiayun Wang, Yubei Chen, Rudrasis Chakraborty, and Stella~X Yu.
\newblock Orthogonal convolutional neural networks.
\newblock In {\em Proceedings of the IEEE/CVF Conference on Computer Vision and
  Pattern Recognition}, pages 11505--11515, 2020.

\bibitem{medmnist}
Jiancheng Yang, Rui Shi, and Bingbing Ni.
\newblock Medmnist classification decathlon: A lightweight automl benchmark for
  medical image analysis.
\newblock {\em arXiv preprint arXiv:2010.14925}, 2020.

\bibitem{Zhai}
Xiaohua Zhai, Alexander Kolesnikov, Neil Houlsby, and Lucas Beyer.
\newblock Scaling vision transformers.
\newblock {\em arxiv:2106.04560}, 2021.

\end{thebibliography}
\end{document}